\documentclass[12pt]{article}
\usepackage{bm} 
\usepackage{amssymb}
\usepackage{natbib}
\usepackage{graphicx}
\usepackage{color}
\usepackage{citesort}

\usepackage{amsmath}

\newcommand{\G}{{\cal G}}
\newcommand{\x}{{\bm x}}

\newcommand{\la}{\langle}
\newcommand{\ra}{\rangle}

\newcommand{\ee}{\end{equation}}
\newcommand{\be}{\begin{equation}}

\begin{document}
\bibliographystyle{jfm}
\title{Mesoscopic modeling of a two-phase flow
 in the presence of boundaries: the  Contact Angle}

\author{R. Benzi$^1$, L. Biferale$^{1}$, M.  Sbragaglia$^{1,2}$, S.
  Succi$^{3}$ and F. Toschi$^{3,4}$\\
$^1$ Dipartimento di Fisica, Universit\`a ``Tor
  Vergata'', and\\ INFN, Via della Ricerca Scientifica 1, I-00133
  Roma, Italy.\\ $^2$ Department of Applied Physics, University of
  Twente,\\ P.O. Box 217, 7500 AE Enschede, The Netherlands.
  \\ $^3$IAC-CNR, Viale del Policlinico 137, I-00161 Roma,
  Italy.\\ $^4$INFN, Sezione di Ferrara, via G. Saragat 1, I-44100
  Ferrara, Italy.  }

\maketitle

\begin{abstract}
  We present a mesoscopic model, based on the Boltzmann Equation, for
  the interaction between a solid wall and a non-ideal fluid.  We
  present an analytic derivation of the contact angle in terms of the
  surface tension between the liquid-gas, the liquid-solid and the
  gas-solid phases. We study the dependency of the contact angle on
  the two free parameters of the model, which determine the
  interaction between the fluid and the boundaries, i.e. the
  equivalent of the wall density and of the wall-fluid potential in
  Molecular Dynamics studies.

  We  compare the analytical results obtained in the
  hydrodynamical limit for the density profile and for the surface
  tension expression with the numerical simulations. We compare  also
  our two-phase approach with some exact results obtained by
  \cite{stone} and \cite{Phi1,Phi2} for a pure hydrodynamical
  incompressible fluid based on Navier-Stokes equations with boundary
  conditions made up of alternating slip and no-slip strips.  Finally,
  we show how to overcome some theoretical limitations connected with
  the discretized Boltzmann scheme proposed by \cite{sc} and we
  discuss the equivalence between the surface tension defined in terms
  of the mechanical equilibrium and in terms of the Maxwell
  construction.
\end{abstract}
\section{Introduction}
The physics of molecular interactions at fluid-solid interfaces is a
very active research area with a significant impact on many emerging
applications in material science, chemistry, micro/nano\-engineering,
biology and medicine, see \cite{white,micro,MEMS}. Many problems
require the spreading of a liquid on a solid that may either be a
simple flat and clean surface or present some degree of roughness
contaminated by compounds with different chemical-physical qualities
(\cite{degennes}).  As for most problems connected with surface
effects, fluid-solid interactions become particularly important for
micro- and nano-devices, whose physical behavior is largely affected
by high surface/volume ratios (\cite{tabebook,beskok}) whose direct
consequence is the enhancement of capillary phenomena
 with respect to bulk properties (\cite{degennes,rw}). On the
theoretical side, very little is known because of the difficulties to
match the classical infinite-volume thermodynamics description with
surface effects. From the experimental side
 the
study of the surface properties for the flow-solid interactions is
much more difficult than  the solid-vacuum case (\cite{bico,craig,ou,tabe,onda,pit,vino,zhu1,zhu2,cheng,choi}).

In most cases, to reach quantitative results on specific problems, one
is forced to rely on numerical simulations, especially in presence of
complex boundary conditions. To date, two major approaches dominate
this field from the numerical side.  The first one is based on a pure
hydrodynamical description, with the interaction between the flow and
the solid fully renormalized in terms of {\it ad hoc} boundary
conditions for the hydrodynamical fields
(\cite{barrat,Phi1,Phi2,MD8,stone}). The main drawback is represented
by the difficulty to describe a variety of different solid properties,
with complex roughness landscape and chemical-physical attributes. The
main advantage is that one can directly focus on spatial and frequency
variations up to the typical hydrodynamical scales.

The second approach attacks the problem from an atomistic description,
by integrating the Newton equations for a set of molecules interacting
via a Lennard-Jones potential. This is the basic idea behind Molecular
Dynamics (MD), which requires an additional {\it ad hoc} tuning of the
free parameters entering  the potential between liquid-liquid
molecules and between liquid-solid molecules
(\cite{MD5,MD6,MD8,MD7,MD4,MD3}). These parameters are fine tuned by
comparison with the experiments and are mainly of two types: the
overall strength of the interactions and the typical interacting
distance (fixed by the relative weight between the attractive and
repulsive terms). The main drawback is here represented by the
congenital scale separation between this method and continuum
phenomena (see \cite{brenner}) and consequently the inability to
describe spatial fluctuations on scales which are much larger than the
inter-molecular interactions and temporal fluctuations larger than a
few milliseconds. The advantage is given by its apparent "ab-initio''
nature, although to be of any practical use, the method needs to be
supplemented with experimental data (\cite{KOPLIK,review2}).

In this paper we follow a third route, focusing on a mesoscopic
modeling of the solid-liquid interaction based on Kinetic Theory of
dense fluids (\cite{cerci2,chapman}). 

The method, known as Lattice Boltzmann equation (LBE), directly
accesses spatial and temporal fluctuations at the hydrodynamical level
with the extra bonus of a large flexibility in the description of the
chemical and physical properties of the boundary conditions (see
\cite{zanetti,HJ,HSB} and for exhaustive reviews see
\cite{chendol,BSV}).  With respect to the MD approach, one pays the
price to move the matching with the experimental data to the level of
bulk interaction between Boltzmann distribution function (the
equivalent of liquid-liquid MD potential) and to the boundary
conditions imposed on the Boltzmann equation (the equivalent of the
solid-liquid potential in the MD).  Ideally one would supply the
infinite BBGKY hierarchy (\cite{bbgky,kirkwood}) typical of any
kinetic description, with atomistic information, thereby closing the
problem without any approximation. In most cases, a more practical
approach is taken: in order to derive useful kinetic description, some
educated guess on the many-body BBGKY hierarchy are proposed and
tested a-posteriori.

In this paper, we shall focus on surface effects in presence of phase
coexistence between a liquid and its saturated vapor. In particular,
we aim at investigating how to develop an effective mesoscopic
description of the surface tension between the liquid-gas,
$\gamma_{lg}$, the liquid-solid, $\gamma_{ls}$ the gas-solid
$\gamma_{gs}$ phases and, more practically of the contact angle
$\theta$ (\cite{rw,degennes}) that can be defined from the above
surface tensions: \be \cos\left(\theta\right) =
\frac{\gamma_{gs}-\gamma_{ls}}{\gamma_{lg}}.\ee We will perform an
analytical and numerical study within the {\it mean field} method
proposed by \cite{BE2phase1,sc}, based on a Lattice Boltzmann equation
with an effective {\it two-body potential} described only in terms of
local, single molecule, properties of the fluid (see next section for
a detailed description of the method). The model provides, to our
opinion, the simplest coherent description of the many-body
interaction typical of dense fluids within the Lattice Boltzmann
Equation framework for non-ideal fluids
(\cite{BE2phase1,sc,BE2phase3,Doolen,KWOK2}).

In this paper, we review first the method as defined by
\cite{BE2phase1,sc} for bulk flows (no boundaries) and we extend it to
include the interaction with a given solid surface by the introduction
of suitable boundary conditions. This defines a theoretical scheme
able to incorporate non-ideal effects (phase transitions) triggered by
the presence of the solid boundary and complex fluid properties
connected to the actual density profiles (contact lines, contact
angle, capillary phenomena, surface tensions, etc...). The main result
of the first part is an {\it exact analytical expression} of the
contact angle in terms of the surface tensions derived from the
hydrodynamical limit of the Boltzmann equations.  In the second part,
we perform a systematic study of the contact angle dependency on the
boundary properties and we compare our two-phase approach with some
exact results obtained for a pure hydrodynamical single-phase fluid
based on Navier-Stokes equations with suitable boundary conditions
(\cite{stone,Phi1,Phi2}).  We also show how to overcome some
theoretical limitations connected with the discretized Boltzmann
scheme here utilized, proposed in \cite{sc}, and we discuss the
equivalence between the surface tension defined in terms of the
mechanical equilibrium or in terms of the thermodynamical ``Maxwell
construction'' (\cite{Stanley,huang}). Finally, we discuss the
possible application of this method to describe non-stationary flows
in micro-channel, the apparent slip phenomenon and the
wetting/dewetting transition induced by micro-corrugation in the
boundaries.

\section{The Shan-Chen (SC)  approach to non-ideal fluids: 
the inclusion of boundaries.}

As soon as one goes beyond the ``ideal gas'' description, allowing
for phase transition and for density and temperature variations inside
the flow, the correct way to approach the kinetic problem is to start
from the BBGKY formalism (\cite{kirkwood}). Phase transitions are
triggered by critical dependency of the thermodynamic variables on
small variations in the local density, temperature and pressure
fields. In order to describe such phenomena, one needs to go beyond
the description based on the probability density to observe a single
molecule with a given velocity, ${\bm v}_{1}$, at position ${\bm
  r}_{1}$ and at time $t$, $f_1({\bm r}_{1}, {\bm v}_{1},t)$.  In
particular, one needs to consider at least the two-particle
distribution, $f_{1,2}({\bm r_1}, {\bm v_1},{\bm r_2}, {\bm v_2},t)$,
which explicitly enters in the Boltzmann equation via the collisional
term, $\Omega$: \be\label{BBGKY1} \partial_t f_1 + {\bm v}_1\cdot
\partial_{\bm r_1} f_1 + {\bm K}_1\cdot \partial_{\bm v_1} f_1 =
\Omega \ee where ${\bf K}_1$ is an external body force and \be \Omega
= -\int d{\bm v}_2 d {\bm r}_2 \, \partial_{{\bm v}_1}f_{1,2} \,
\partial_{{\bm r}_1}\, V(r_{12}) \ee with $ V(r_{12})=V(|{\bm
  r}_1-{\bm r}_2|)$ being the inter-particle potential.  The BBGKY
hierarchy prescribes the evolution of $f_{1,2}$ in terms of the three
particle densities, $f_{1,2,3}$, the evolution of $f_{1,2,3}$ in terms
of the four-particles density and so on. The simplest closure which
takes into account the two-particles interaction consists in adopting
a ``mean field'' approach for the collisional terms. This approach
starts by rewriting $f_{1,2}$ in the equivalent form:
$$f_{1,2}({\bm r_1}, {\bm v_1},{\bm r_2}, {\bm v_2}) = f_1({\bm r_1},
{\bm v}_1)f_2({\bm r_2}, {\bm v}_2) g({\bm r}_1,{\bm r}_2,{\bm
  v}_1,{\bm v}_2)$$ where we have introduced the two particles
correlation function, $g({\bm r}_1,{\bm r}_2,{\bm v}_1,{\bm v}_2)$.
To proceed further one needs to make some approximation on the
two-body correlation function $g({\bm r}_1,{\bm r}_2,{\bm v}_1,{\bm
  v}_2)$. The celebrated ``molecular chaos'' assumption of Boltzmann
gives $g =1$, i.e. absence of both velocity and spatial correlations
(\cite{cerci2}). In the less restrictive case where  only 
velocity correlations vanish, one has:
$$ g({\bm r}_1,{\bm r}_2,{\bm v}_1,{\bm v}_2) = g({\bm r}_1,{\bm r}_2)
$$ and the collisional term can be (see also \cite{Martys}) rewritten
as: \be
\label{eq:twoparticles}
\Omega =-\partial_{{\bm v}_1}f_{1} \int d{\bm r}_2 \rho({\bm r_2})
g({\bm r}_1,{\bm r}_2) \partial_{{\bm r}_1} V(r_{12}) \ee where we
have used the definition of the local density as
\be\label{localdensity} \rho({\bm r},t) = \int d{\bm v} f({\bm r},{\bm
  v},t).  \ee The approximation (\ref{eq:twoparticles}) is at the core
of many Lattice Boltzmann description of non-ideal fluids because now
the collisional term has the form of a body force term, $\Omega
\propto {\bm K}_1 \cdot \partial_{\bm v_1}f_1$, and may be seen as a
renormalization of the local pressure tensor via the introduction of
non-ideal terms in the equation of state
(\cite{chendol,heshan}).

More quantitatively, if we start from equation (\ref{BBGKY1}) and
together with (\ref{localdensity}) we define the local momentum as

\be\label{momentum} \rho {\bm u}({\bm r},t)=\displaystyle \int d{\bm
  v} f({\bm r},{\bm v},t) {\bm v} \ee we obtain (see
\cite{Martys,kirkwood}) conservative equations for the two local
fields: \be\label{integralden} \partial_{t} \rho + {\bm \nabla} \cdot
(\rho {\bm u})=0 \ee \be \label{integralmom} \partial_{t} (\rho {\bm
  u}) + {\bm \nabla} \cdot ( \int d{\bm v} f {\bm v} {\bm v}) =
\displaystyle\int d{\bm v} \Omega {\bm v} = -{\bm \nabla}
\overleftrightarrow{W} \ee where the tensor $\overleftrightarrow{W}$
is directly related to the interaction potential $V(r_{12})$: \be
W_{i,j}=-\frac{1}{2}\displaystyle\int d {\bm r_1} d{\bm v_{1}} d {\bm
  v_{2}} d{\bm r_{2}} \, f_1\, f_2\, g \,\delta({\bm r}- {\bm r_1})\,
V^{\prime}(r_{12}) r^{-1}_{12}(r_{12})_{i}(r_{12})_{j}.  \ee The
previous equations are clearly locally conservative for density and
globally conservative for momentum (\cite{kirkwood}).

In the realm of Boltzmann Equations, a popular way to simplify the
collisional integral is to write it as a simple relaxation term (with characteristic time $\tau$) towards
a suitable local equilibrium.  This is the celebrated BGK
approximation given by \cite{BGK} and one may wonder which is the simplest BGK
description consistent with equations
(\ref{integralden},\ref{integralmom}). By consistent BGK description,
we mean a single time relaxation collisional term of the form: \be
\Omega_{BGK}=-\frac{1}{\tau}\left(f-f^{(M)}\right) \ee where the
equilibrium distribution $f^{(M)}$ is the local maxwellian in $D$
dimensions: \be\label{MAX} f^{(M)}=\frac{\rho^{\prime}}{(2 \pi K
  T^{\prime})^{D/2}}\exp\left\{\frac{({\bm v}-{\bm
    u^{\prime}})^2}{2KT^{\prime}}\right\} \ee and the parameters
$\rho^{\prime}({\bm r},t)$, ${\bm u^{\prime}}({\bm r},t)$,
$T^{\prime}({\bm r},t)$ can be chosen in such a way to be consistent
with global balance equations.  In particular straightforward Gaussian
integration yields to $\rho^{\prime}({\bm r},t)=\rho({\bm r},t)$ and
\be {\bm u}^{\prime}({\bm r},t)={\bm u}({\bm r},t)-\frac{1}{\tau \rho}
    {\bm \nabla} \overleftrightarrow{W} ({\bm r},t) \ee which means a
    space-time dependent shift of the mean-value of the local
    Maxwellian. The Shan-Chen model (\cite{BE2phase1,sc}) is precisely
    equivalent to such an approach with the assumption 

\be {\bm \nabla} \overleftrightarrow{W} ({\bm r},t) = \displaystyle
    {\cal G}_{b} \int_{0}^{s_{max}} ds \int d \Omega_v \,\psi({\bm r},t)
    \psi ({\bm r}+{\hat {\bm v}}s,t) {\hat{ \bm v}}
\label{eq:inter}
\ee
where with $d \Omega_v$ we denote the integration over the
 angular dependency of the unit vector, $\hat{\bm v}$ and  
 where $\psi({\bm r},t)$ is a mean field potential which  depends on
${\bm r}$ and $t$ only through the local density, i.e. $\psi({\bm
  r},t) \equiv \psi(\rho({\bm r},t))$. In the above, $|s_{max}
 \hat{\bm  v}|$ represents the range of the interactions and ${\cal{G}}_{b}$ a
coupling constant, something like an inverse temperature for the
model.

In principle, the temperature $T^{\prime}({\bm r},t)$ in (\ref{MAX})
should also be changed on account of thermodynamics consistency and
total energy conserving dynamics (potential plus kinetic).  However,
being interested in isothermal phenomena, we keep
$KT^{\prime}=c^2_{s}$, $c^{2}_{s}$ being the sound speed velocity (for
a possible extension of the model to include also temperature
fluctuations see \cite{Martys}).

Upon discretization of the approximation described before, we derive
immediately the (Lattice) Boltzmann Equations
(\cite{Saurobook,gladrow}) as follows: \be
\label{eq:lbe}
f_\alpha(\x+ {\bm c}_{i} \Delta t,t+\Delta t)-f_\alpha(\x,t)=-\frac{
  \Delta t}{\tau} \left[f_\alpha(\x,t)
  -f_\alpha^{(eq)}(\rho(\x,t),{\bm u}^{\prime}(\x,t))\right] \ee where
${\bm x}$ runs on a two (or three) dimensional Lattice and $\Delta t
=1$ is the time stepping in the numerical scheme. The LHS of
(\ref{eq:lbe}) is the molecular free streeming of a discrete set
(${\bm c}_{\alpha}, \alpha=0,...,N$) of velocities whereas the
right-hand side represents molecular collisions via a simple
relaxation towards the local equilibrium $f_\alpha^{(eq)}$ (the  local
Maxwellian expanded to second order in the Mach number) in a time
lapse of the order of $\tau$.  This relaxation time fixes the fluid
kinematic viscosity as $\nu=c_s^2(\tau-1/2)$ 
where $c_s = 1/\sqrt{3}$ in the present work (see \cite{gladrow}).
  The fluid density and
momentum are given by
$$\rho=\sum_\alpha f_\alpha \; \;\;\;\;\;\;\; {\bm u} =
\frac{1}{\rho}\sum_\alpha f_\alpha {\bm c}_\alpha$$ and can be shown
to evolve according to the Navier-Stokes equations of fluid-dynamics,
\cite{gladrow}: \be \left \{ \begin{array} {l} \rho [ \partial_{t}
    {\bm u} + ({\bm u} \cdot {\bm \nabla} ){\bm u}] = - {\bf \nabla}
  \overleftrightarrow{P_0} +{\bm F} + {\bm \nabla} \cdot (\nu \rho
                     {\bm \nabla} {\bm u})\\ \partial_{t}\rho + {\bm
                       \nabla} \cdot (\rho {\bm u}) = 0 \end{array}
\right.
\label{eq:NS}
\ee 
being $ \overleftrightarrow{P_0}$ the ideal pressure tensor given by
the perfect-gas equation of state: $
\overleftrightarrow{P_0}=\overleftrightarrow{I} c^{2}_{s} \rho$, where
$\overleftrightarrow{I}$ is the unit tensor. Non-ideal effects are
modeled through the self-consistent body force term (\ref{eq:inter})
that can be discretized as: 
\be {\bm F}(\x,t) = -\G_{b}
\psi(\x,t)\sum_\alpha w_{\alpha} \psi(\x+{\bm c_{\alpha}} \Delta
t,t){\bm c}_{\alpha} \ee 
where $\psi(\x,t)$ is the lattice version of the mean field potential
previously used and $w_{\alpha}$ are normalization weights (see
Appendix A for more technical details). Due to this body force at each
time step we consistently re-define the velocity ${\bm u}^{\prime}$ in
the equilibrium distribution $f^{(eq)}$ as:
$$ {\bm u}^{\prime} = \frac{1}{\rho} \sum_\alpha {\bm c}_\alpha
f_\alpha -\frac{1}{\tau \rho} \G_b \psi(\x,t)\sum_\alpha \psi(\x+{\bm
  c}_\alpha\Delta t,t){\bm c}_\alpha
$$ 
which in turns implies a non-diagonal pressure tensor
$\overleftrightarrow{P}$ deduced from the condition:
\be -{\bm \nabla} \overleftrightarrow{P} -{\bm \nabla}
\overleftrightarrow{P_0} = {\bm F}. \label{eq:P} \ee 
By expanding in Taylor series the inter-particle potential one gets
(see Appendix A) in the case $\Delta t =1$:
\be
\label{TENSORE}
 P_{ij}=\left[
   c^{2}_{s}\rho+\frac{1}{2}c^{2}_{s}\G_{b}\psi^{2}+\frac{1}{2}c^{4}_{s}\G_{b}\psi
   \Delta \psi + \frac{\G_{b} c^{4}_{s}}{4} |{\bm \nabla} \psi|^{2}
   \right] \delta_{ij}-\frac{1}{2}c^{4}_{s}\G_{b} \partial_{i} \psi
   \partial_{j}\psi.  \ee Let us notice that there is always a certain
   degree of arbitrariness in deriving the full expression of the
   pressure tensor in the continuum limit \cite{rw}. 
The most reasonable way to do it is to impose that the constraint
   (\ref{eq:P}) is verified up to the second order in the Taylor
   expansion of both pressure and forcing expression. We stop at
   second order because Navier-Stokes equations are obtained from the
   LBE at the second order in the Chapman Enskog expansion. 
The above expression for the pressure tensor is
   different from the one proposed in equation (19) of \cite{sc}. The reason
   is that in \cite{sc} the constraint (\ref{eq:P}) is verified only
   up to the first order in Taylor expansion. This difference is
   important for the thermodynamic consistency of isothermal flow 
as it will
   be described in section \ref{sec:thermo}, see also \cite{Doolen}.

 The first two terms in the diagonal part of (\ref{TENSORE})
   describe the bulk homogeneous phase transition by the non-ideal
   equation of state: \be
\label{eq:eqstate}
P_{b}(\rho) = c^{2}_{s}\rho+\frac{1}{2}c^{2}_{s}\G_{b}\psi^{2}(\rho).
\ee with a typically used functional form: \be
\label{eq:psi1}
\psi(\rho)=(1-e^{-\rho/\rho_{0}}) \ee with $\rho_0$ a reference
density.

Now, we will consider a general background that is independent of the
functional forms of $\psi$, given for granted that, for a quantitative
agreement with MD and experiments, one can choose different functional
forms and/or different inter-particle interactions. We will be back on
the importance of the functional form later on in section
(\ref{sec:thermo}), where we investigate the Maxwell construction in
the $(P,V)$ diagram of the model. Let us only discuss for the moment
the qualitative behavior imposed on the $\rho$, dependency of
$\psi(\rho)$ by two physical constraints. First, for small densities
$\rho$ we need to recover the equation of state of an ideal gas, which
requires that:
 $$\psi(\rho) \propto \rho \qquad \rho \rightarrow 0.$$ Second, for
large local densities, the interacting potential must saturate:
$$ \psi(\rho) \rightarrow {\mbox{const}}. \qquad \rho \gg \rho_0$$ a
requirement meant to mimic the hard-core properties of real molecules
which prevent unphysical density accumulations.  Any smooth functional
form which satisfies the two previous requirements leads to a phase
transition as soon as $\G_{b}$ becomes smaller than a critical value
$\G_c$ (with $\G_c$ being negative), 
where the fluid start to exhibit two coexisting phases with
the same pressure. All the other terms in eq.~(\ref{TENSORE}), which
depend explicitly on the density variations, describe the development
of an interface profile with its own surface tension (as soon as the
isotropy of the pressure tensor is violated).

\subsection{SC model without boundaries}
For the sake of completeness, let us summarize again the steps needed
to calculate the density profile in presence of a liquid-gas interface
in an unbounded domain as shown for the first time by \cite{sc}. This
calculation will be used later on to implement the expression of the
contact angle in presence of boundaries.\\
\noindent In order to calculate the density profile we need to use the general
expression of the pressure tensor (\ref{TENSORE}) and insert it in to
the mechanical equilibrium condition: 
\be
\label{nabla}
{\bm \nabla} \overleftrightarrow{P} ({\bm x})=0,  \ee 
with the appropriate boundary conditions $\rho(-\infty)=\rho_g$ and
$\rho(+\infty)=\rho_l$ for a planar interface between liquid and
gas. Let us suppose that the interface develops along the $y$
coordinate. Under this geometry, the pressure tensor becomes
anisotropic with a mismatch between the transverse components,
$P_{xx}(y)=P_{zz}(y)$ and the normal component $P_{yy}(y)$. The
condition that the interface does not move, (\ref{nabla}), implies
that the normal component remains constant and equal to the value in
the bulk, $P_{bulk}$, throughout the interface: 
\be P_{yy}(y)\equiv P_{bulk}=c^{2}_{s}\rho+\frac{1}{2}c^{2}_{s}\G_{b}
\psi^{2}+\frac{1}{2}c^{4}_{s}\G_{b}\psi \partial_{yy} \psi -
\frac{\G_{b} c^{4}_{s}}{4} |\partial_{y} \psi|^{2}.
\label{eq:density}
\ee 
The density shape is now fully determined by solving
(\ref{eq:density}) with the requirement that the liquid and gas phase
share the same value of the bulk pressure: \be
P_{bulk}=c^{2}_{s}\rho_g+\frac{1}{2}c^{2}_{s}\G_{b}\psi^{2}(\rho_g) =
c^{2}_{s}\rho_l+\frac{1}{2}c^{2}_{s}\G_{b}\psi^{2}(\rho_l).
\label{eq:state1}
 \ee 
By making the change of variables $(d\rho/dy)^2 = z$, one may rewrite
the mechanical equilibrium as an ordinary differential equation for
the inter-particle potential where only derivatives with respect to
the density $\rho$ appear: \be P_{bulk}=c^{2}_{s} \rho
+\frac{\G_{b}}{2}c^{2}_{s} \psi^{2} +\frac{\G_{b}c^{4}_{s}}{2} \left[
  \frac{1}{2}\frac{\psi^2}{\psi^{\prime}} \frac{d}{d \rho} \left(
  \left( \frac{d \rho}{d y} \right)^2 \left( \frac{d \psi}{d \rho}
  \right)^2 \frac{1}{\psi} \right) \right]
\label{eq:ode} .
\ee
The above differential equation can be integrate explicitly to give:
\be
\label{eq:ode2}
z(\rho) = \frac{4 \psi}{\G_{b} c^{4}_{s}(\psi^{\prime})^2}
\displaystyle\int_{\rho_g}^{\rho}\left(P_{bulk}-c^{2}_s\rho-\frac{\G_{b}}{2}c^{2}_{s}\psi^2
\right)\frac{\psi^{\prime}}{\psi^2}d \rho \ee 
with $\psi^{\prime}=\partial \psi/\partial \rho$. If the two extremes
of the integral are chosen inside the bulk phases we get: $z(\rho) =
0$ for $\rho=\rho_l$ and $\rho=\rho_g$. It is easy to realize that in
order to be compatible with the latter boundary conditions, we must
require: 
\be
\label{eq:inte}
\int_{\rho_l}^{\rho_g} \left[P_{bulk}-c_s^2 \rho
  -\frac{c_s^2\G_b}{2}\psi(\rho)^2\right] \frac{\psi'}{\psi^2} d \rho = 0
\ee 
which fixes, together with (\ref{eq:state1}), the two densities
$\rho_l$ and $\rho_g$ as shown by \cite{sc}.

The whole profile can be obtained by inverting (\ref{eq:ode2}) or by
directly numerically solving (\ref{eq:ode}) starting from the inside
of one bulk phase and making a small initial spatial perturbation on
the constant density profile.  Let us notice that equations
(\ref{eq:density},\ref{eq:ode2},\ref{eq:inte}) are different from
equations (21,24,25) of \cite{sc} because of the different requirements imposed here 
in the derivation of the pressure tensor as discussed after equation 
(\ref{TENSORE}).
\subsection{Surface tension and Contact Angle}

Following \cite{rw}, we define the Liquid-Gas surface tension,
$\gamma_{lg}$, as the integral along the coordinate normal to the
interface of the mismatch between normal and transverse components of
the above tensor. More precisely, assuming that the only dependence is
in the $y$ coordinate, the local increment of the surface tension is:
\be
\label{gainsurf} 
P_{yy}-P_{xx} = \frac{d \, \gamma_{lg}}{d y}=-\frac{1}{2}c^4_s
\G_b|\partial_{y} \psi|^2.  \ee
Upon integration from $-\infty$ to $+\infty$, the surface tension is
readily calculated:
\be
\label{tensione}
\gamma_{lg}= -\frac{1}{2}c^{4}_{s}\G_{b}
\displaystyle\int_{-\infty}^{+\infty}|\partial_{y}\psi|^{2} dy.  \ee
This implies the existence of transverse stresses that would result in
pressure drop for spherical interfaces and in the most general Laplace
law for a curved surface: \be \Delta p = \gamma_{lg} \left(
\frac{1}{R_1}+\frac{1}{R_{2}} \right) \ee being $R_1$ and $R_2$ the
local principal radii of curvature of the surface, see
\cite{rw,degennes,sc}.  Let us now discuss how to generalize the
previous results to the case of a solid boundary. In our language,
mechanical equilibrium between multi-phase systems (say liquid, gas
and solid) can be formulated as a more general problem. We want to
study the mechanical equilibrium by imposing that the density profile
must match some given value at the boundary position: 

\be
\label{MASTER}
\left \{ \begin{array} {l l} 
{\bm \nabla} \overleftrightarrow{P} ({\bm x})=0 \\
\psi(\rho({\bm x}_{w}))=\psi(\rho_w)\\
%\psi(\rho(+\infty))=\psi(\rho_{\alpha})\\
\end{array} \right.
\ee 
where the solid wall is at position $\x_w$. Equation (\ref{MASTER}) is
nothing but the mechanical equilibrium of a multiphase system in the
presence of a boundary condition. The value of the inter-particle
potential at the wall, $\psi(\rho_w)$, is a free parameter in the
model, and it is not meant to be related with the ``true'' density of
the solid phase. It will be used to tune different wall
properties.

Focusing, for the sake of simplicity, on two-dimensional systems (see
figure \ref{fig:angle} in Appendix B) the mechanical equilibrium
equation translates into \be \partial_{x} P_{xx}+\partial_{y} P_{xy}=0
\ee or equivalently ($\partial_{x} P_{xy}+\partial_{y} P_{yy}=0$),
whose meaning is that the flux over an arbitrary contour (Green's
theorem) of the vector $(P_{xx},P_{xy})$ is zero. With reference to
figure \ref{fig:angle} in Appendix B we notice that for such a
calculation we need to specify the pressure tensor along a solid-gas
interface and liquid-solid interface.  If we choose the rectangular
contour shown in figure \ref{fig:angle} and impose the flux of the
above vector exactly to zero (details are given in the Appendix B) we
obtain:

\be
\label{eq:ca}
 \cos{(\theta)}=\frac{\displaystyle\int_{sg} |\partial_{y} \psi|^2
   dy-\displaystyle\int_{sl} |\partial_{y} \psi|^2
   dy}{\displaystyle\int_{lg} |\partial_{y} \psi|^2 dy} \ee 
where $\displaystyle\int_{sl} |\partial_{y} \psi|^2 dy$,
$\displaystyle\int_{sg} |\partial_{y} \psi|^2 dy$,
$\displaystyle\int_{lg} |\partial_{y} \psi|^2 dy$ indicate the
positive integrals calculated along the solid-liquid, solid-gas and
liquid-gas interfaces.

From the above relation (\ref{eq:ca}), which fully determines the
contact angle in our LBE scheme, one naturally extracts the definition
of the surface tensions: 
\be
\label{eq:sf3}
\gamma_{\alpha,\beta} =  -\frac{1}{2}c^{4}_{s}\G_{b} \displaystyle\int_{\alpha,\beta} |\partial_{y}
\psi|^2dy \ee 
where with $\alpha,\beta$ we mean any two among the liquid, gas and
solid phases. Notice that, rigorously speaking, the surface tensions
between liquid-solid and between gas-solid are defined only modulo an
additive constant, being operationally defined in terms of the contact
angle which depends only on their difference (\cite{degennes}).  The
above definition is consistent with the requirement to have
$\gamma_{ls}=0$ when the wall has the same density of the liquid
(perfect wetting).  Correspondingly, one may also imagine a situation
when the gas phase perfectly matches the wall properties,
$\rho_g=\rho_w$, where we have $\gamma_{sg}=0$ and consequently a
complete dewetting.

\subsection{Analytical and Numerical results}

\begin{figure}
\begin{center}
\includegraphics[scale=.75]{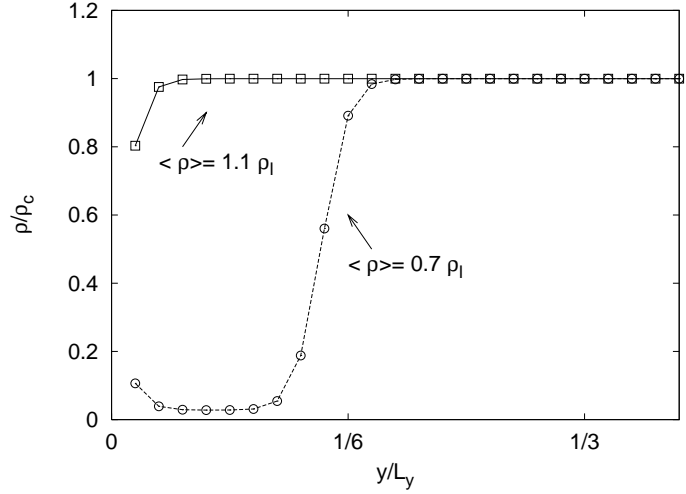}
\caption{The stationary state configuration for the density field as a
  function of the relative distance from the wall. We present the
  density field normalized to the center channel density for two
  different values of the average density of the system: $\la \rho
  \ra= 1.1 \rho_{l}$ ($\square$) and $\la \rho \ra = 0.7 \rho_{l}$
  ($\circ$). In both cases we integrate numerically the Lattice
  Boltzmann Equation with $\tau=1$ in a $2d$ channel with two walls
  (bottom and lower) and periodic boundary condition in the stream-wise
  direction. The dimensions of the channel are $L_{x}\times
  L_{y}=80\times 45$ and $\G_{b}=-6.0$, being $\G_{c}=-4.0$ the
  critical point of the system. The liquid and gas density for this
  value are respectively $\rho_{l}=2.65$ and $\rho_{g}=0.07$ in
  Lattice Boltzmann units. Notice that the density tends to match a
  given value at the wall (for $y/L_{y}=0$) which is different from
  both liquid and gas values, this is because of the chosen boundary
  condition, $\psi(\rho_w=0.5)$.}
\label{fig:density}
\end{center}
\end{figure}

From the mechanical equilibrium condition (\ref{MASTER}), we may
calculate the density variation along an interface between any two of
the three phases, generalizing the calculation made by \cite{sc} for
the gas-liquid interface only. To accomplish this, we must integrate
the equation for the normal component of the pressure tensor
(\ref{eq:density}) imposing the boundary conditions at the solid,
$\rho(y=y_w)=\rho_{w}$, and $\rho(\infty)=\rho_{\beta}$ where with
$\beta$ we denote either the liquid or the gas phase (see also section
\ref{sec:thermo} for more details). In figure \ref{fig:density} we
show two such profiles.  In our first case at the center of the
channel there is only liquid and the gas phase cannot develop,
i.e. the averaged density is larger than the liquid density at
coexistence, while in the second case the averaged density is chosen
such that a gas phase can develop between the liquid and the wall.

\begin{figure}
\begin{center}
  \includegraphics[scale=.75]{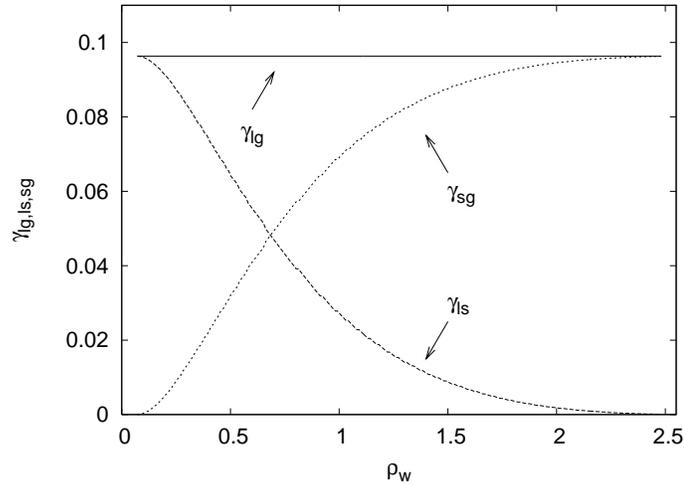}
  \caption{Surface tensions in a $3$ phase system for $\G_b=-6.0$ as a
    function of $\rho_w$.  The horizontal line represents the surface
    tension between liquid and gas ($\gamma_{lg}$). The surface
    tension between solid and gas ($\gamma_{sg}$) is zero for
    $\rho_w=\rho_{g}=0.07$ and equal to $\gamma_{lg}$ for $\rho_w=
    \rho_l = 2.65$. The surface tension between liquid and solid
    ($\gamma_{ls}$) reflects the opposite behavior.}
\label{figure6}
\end{center}
\end{figure}

Once the density profile is known , it is easy to obtain the
corresponding surface tension by plugging the density profile into the
expression (\ref{eq:sf3}). For example, in figure \ref{figure6}, we
show the surface tensions, $\gamma_{ls},\gamma_{lg},\gamma_{sg}$ at a
given temperature ($\G_{b}$) and by changing $\rho_w$, i.e. by varying
the wetting properties of the surface. Of course, only the first two
will depend on $\rho_w$ and, accordingly to our previous discussion,
they must satisfy $\gamma_{ls}=0$ for $\rho_w=\rho_l$ (perfect
wetting) and
$\gamma_{gs}=0$ for $\rho_w=\rho_g$.

\begin{figure}
  \begin{center}
    \includegraphics[scale=.75]{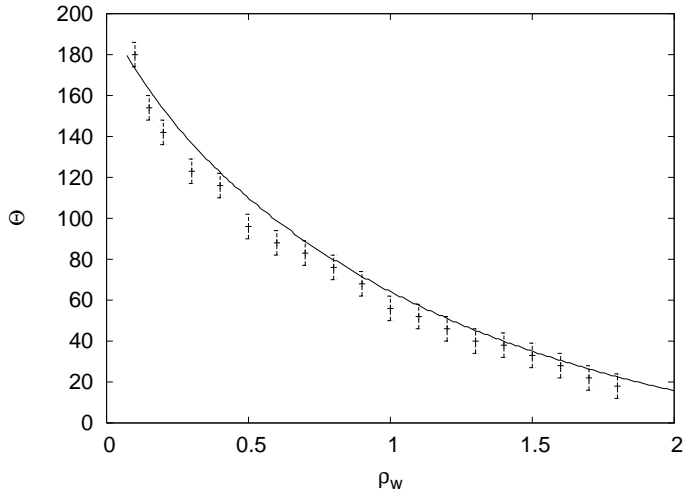}
    \caption{Contact angle for the case $\G_b=-6.0$ as a function of
      $\rho_w$. The analytical estimate is compared with the
      experimental results of Lattice Boltzmann simulations ($+$). The
      Lattice Boltzmann simulations are carried out in a $2d$ domain
      periodic along the stream-wise direction and with two walls
      (upper and lower wall). The grid mesh used is $L_{x}\times
      L_{y}=80\times 45$ and the relaxation time chosen is
      $\tau=1$. To produce the steady state with a drop of liquid the
      system has been initialized with a non-homogeneous condition
      with a square spot of liquid on the lower wall.}
    \label{figcontact}
  \end{center}
\end{figure}

\begin{figure}
  \begin{center}
    \includegraphics[scale=.6]{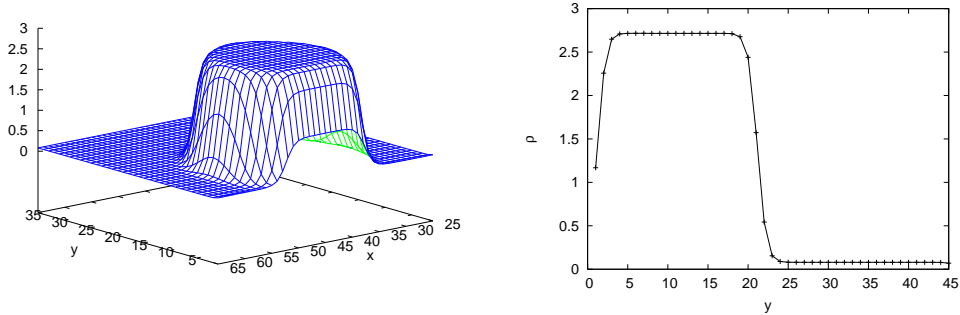}
    \caption{Left: typical  density profile from which
we extract the contact angle. The contact angle in the numeric is extracted by
      estimating the angle made by the surface where the density
      profile is equal to the average value $(\rho_{l}-\rho_{g})/2$
      being $\rho_{l}=2.65$ and $\rho_{g}=0.07$ for the present case
      with $\G_{b}=-6.0$. The computational parameters are the same
      used for figure \ref{figcontact}. To produce the three phases
      equilibrium the system has been initialized to a non-homogeneous
      condition with a a square spot of liquid in contact with the
      lower wall. Right: a vertical snapshot of the left panel for $x=45$.}
    \label{fig:gonio}
  \end{center}
\end{figure}

Given the surface tensions for a fixed temperature one may readily
calculate the contact angle which describes the macroscopic properties
of the surface as a function of the mesoscopic boundary condition
imposed on $\psi({\bm x}_{w}) = \psi(\rho_w)$.
This is the first important methodological result of this paper,
i.e. we provide a mesoscopic way to parametrize the hydrodynamical
behavior of fluid in the proximity of a surface (with different
contact angles) as a function of the tunable free parameter
$\psi(\rho_w)$.\\
\noindent
In figure \ref{figcontact} we show the analytical results obtained by
inserting the density profile from (\ref{eq:ode}) into
(\ref{eq:ca}-\ref{eq:sf3}). In the same figure we also show the
numerical results obtained by running the LBE code and with the
estimate of the contact angle obtained with a goniometer as shown in
figure \ref{fig:gonio}. As one can see from figure \ref{figcontact},
the method is able to reproduce very hydrophobic material, $\theta
\sim 180^o$ and perfectly wetted surface $\theta \sim 0^o$, the latter
case is obtained when $\rho_w$ is chosen equal to the liquid density
at the given temperature. All numerical results are obtained by using
the shape (\ref{eq:psi1}) for the inter-particle potential. This is
the first study, to our knowledge, where a systematic analytical
procedure to derive the contact angle in Lattice Boltzmann models with
interparticle potentials has been proposed. Other important attempts
were already published by \cite{YeomansWag} concerning a lattice
transcription of mean field thermodynamic boundary conditions
(\cite{Cahn}) in the framework of free-energy multiphase methods (see
\cite{BE2phase3}). Numerical investigations of the contact angle
within a LBE approach were also presented by \cite{KWOK,KWOK2} but
without any analytical control on the links with the surface tension
as proposed here.

\begin{figure}
\begin{center}
  \includegraphics[scale=.75]{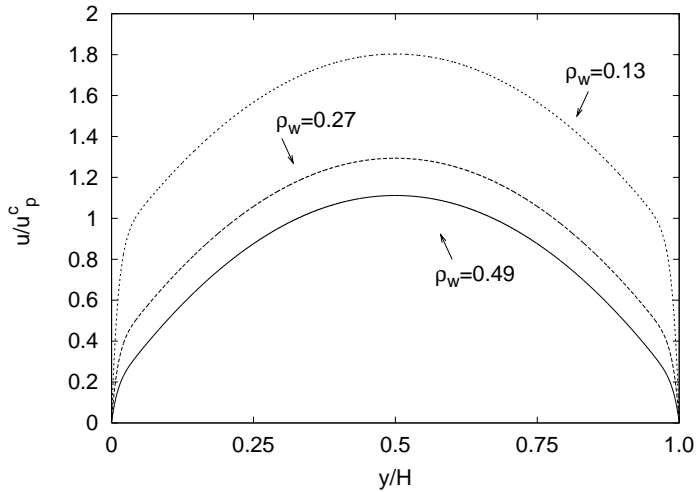}
  \caption{Velocity profiles for pressure driven Stokes flow.
    Profiles are obtained integrating numerically the Stokes equation
    for unit pressure drop and kinematic viscosity (\ref{stokesapp})
 in a homogeneous
    (in the stream wise direction) channel with an height $L=50$ grid
    points. The profiles are then normalized with respect to the
    center channel velocity of the Poiseuille flow ($u^{p}_{c}$) for
    the same geometry.  From top to bottom we show different values of
    $\rho_{w}$ corresponding to different contact angles:
    $\rho_{w}=0.13$ ($\theta=166^{\circ}$), $\rho_{w}=0.27$
    ($\theta=145^{\circ}$), $\rho_{w}=0.49$ ($\theta=110^{\circ}$).}
\label{fig:velocityprofile}
\end{center}
\end{figure}

The production of a rarefaction zone close to the wall is also
important to determine the slippage properties in the case when a
constant external pressure drop is applied on the system (Poiseuille
flow).  For instance, the formation of a gas layer is believed to be
the most probable cause of the {\it apparent} slip length measured in
many experimental micro-channels (see \cite{tretheway}). The idea is
that the rarefaction layer leads to two feedback on the bulk fluid
velocity. First, it allows for the fluid to slide on it without
touching the boundary. Second, it gives an effective reduction on the
channel width seen by the fluid, leading to an overall increase of the
mass throughput for a given pressure drop. Once the density profile is
determined, one may solve the equations for the stream-wise velocity
(\ref{eq:NS}) profile in presence of a unitary external pressure
gradient, $\nabla P_{ext}$, and a unitary kinematic viscosity, $\nu$,
in the Stokes approximation: 
\be\label{stokesapp} \partial_{y} (\rho
\partial_{y} u_{x})= \frac{\nabla P_{ext}}{\nu}= 1.  \ee 
In figure
\ref{fig:velocityprofile} we show a few examples of the velocity
profile at changing the contact angle.  All data have been obtained by
imposing that the flow has the fluid density ($\rho_{l}$) at the
center of the channel and that it ends with a density ($\rho_{w}$) at
the boundary.  Notice that, by increasing the hydrophobic property of
the surface (increasing $\theta$), the rarefaction layer becomes more
and more singular, i.e. the velocity profile develops higher and
higher shear rates at the boundary.

\begin{figure}
  \begin{center}
    \includegraphics[scale=.5]{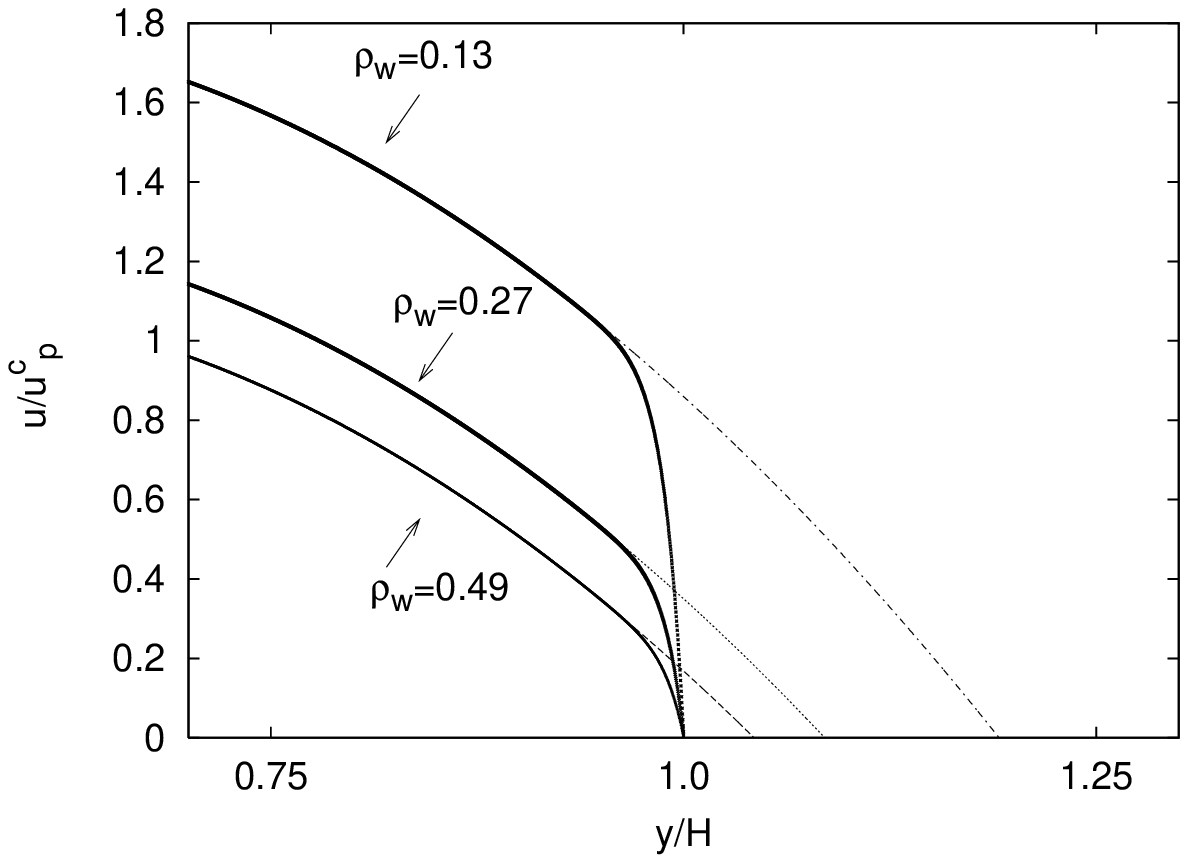}
    \includegraphics[scale=.5]{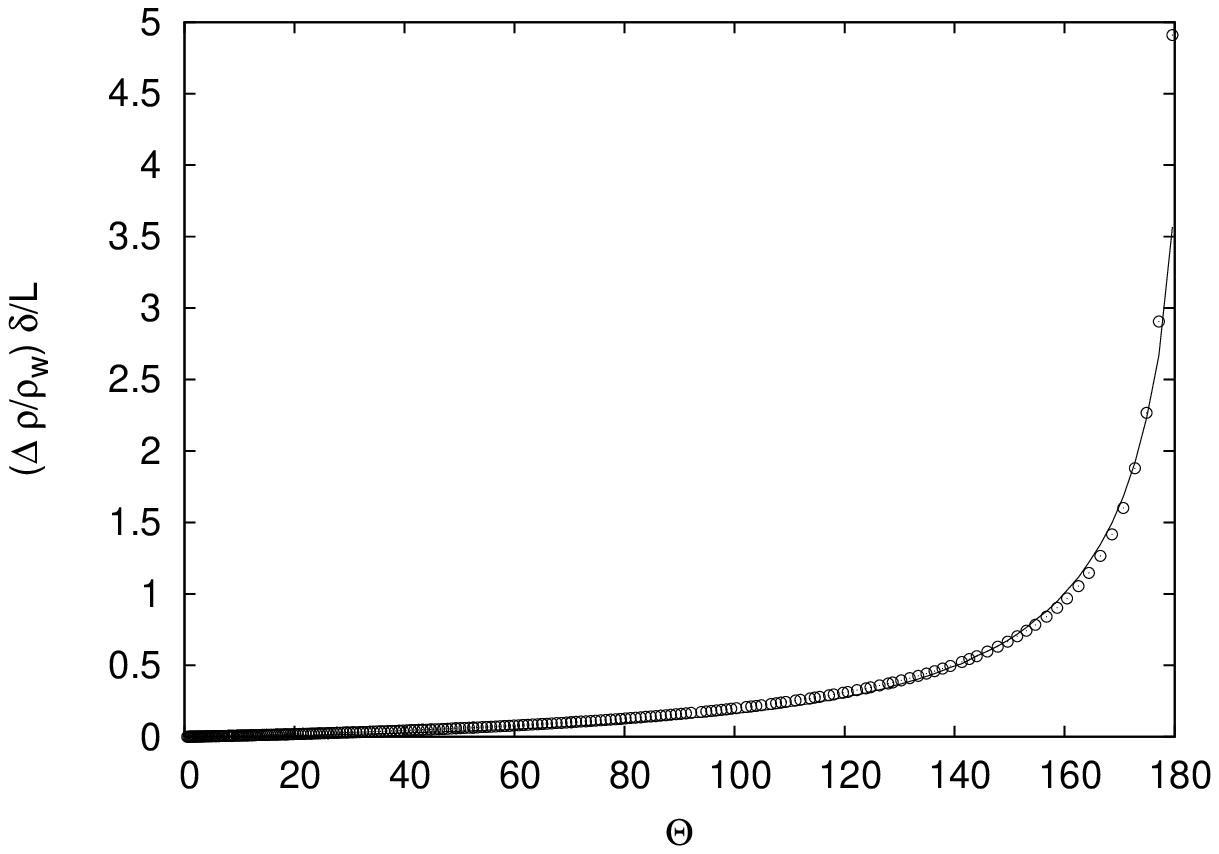}
    \caption{Slip length for the profiles in the Stokes
      approximation. In the left panel we show an extrapolation of the
      profiles showed in the previous figure. The slip length is
      obtained extrapolating to zero the parabolic profile obtained as
      a best fit in the bulk region of the channel. In the right panel
      we show the slip length estimated indirectly from the mass flow
      rate gain with respect to the Poiseuille profile for different
      contact angles ($\circ$) and compare the with the
      phenomenological result (\ref{eq:slip}), solid line.  The good
      agreement is obtained with a prefactor in (\ref{eq:slip}) fixed
      to be $1.2$.}
    \label{fig:slip}
  \end{center}
\end{figure}

To quantify this effect, one usually introduces an {\it apparent slip
  length}, $\lambda_s$, defined as the length were the bulk,
Poiseuille-like, velocity profile extrapolates to zero away from the
wall, see figure \ref{fig:slip}. A simple phenomenological estimate
of $\lambda_s$ for small density variations has been proposed
previously in \cite{PRL} where the authors show that the apparent slip
length can be estimated as: 
\be \lambda_{s} \sim \frac{\Delta \rho}{\rho_w} \delta
\label{eq:slip}
\ee 
where with $\Delta \rho$ we mean the density jump from the bulk to the
wall values and with $\delta$ is proportional to the rarefaction
layer, for a similar study see also \cite{jens}. Another important quantity is the {\it mass flow rate gain},
i.e.  the ratio between the actual mass flow rate and the mass flow
rate obtained assuming a perfect Poiseuille-like profile in the whole
channel section: 
\be
\label{eq:mfr} 
\Phi = \frac{\int \, dy (\rho \, u_x) }{L^3 \nabla P_{ext}/(12 \nu)} \ee 
where $L$ is the total channel height.  The mass flow rate is the only
quantity which can be easily measured in microdevices experiments. It
therefore plays an important role as a benchmark for many modeling
methods. In figure \ref{fig:slip} we show the results for the mass
flow rate gain of our mesoscopic model as a function of the contact
angle. Notice that one can easily gain a factor of the order of $4\div
5$ times larger for the case of high hydrophobic boundaries.  In the
same figure (right panel) 
we also show the phenomenological estimate for the mass
flow rate gain obtained by using the expression (\ref{eq:slip}) with
$\Delta \rho /\rho =(\rho_{l}-\rho_{w})/\rho_{w}$ and $\delta$
estimated as the distance (starting from the wall) needed to reach the
$98 \%$ of the center channel density (again $\rho_{l}$).

\subsection{Consistency with Thermodynamics}
\label{sec:thermo}
\begin{figure}
  \begin{center}
    \includegraphics[scale=.75]{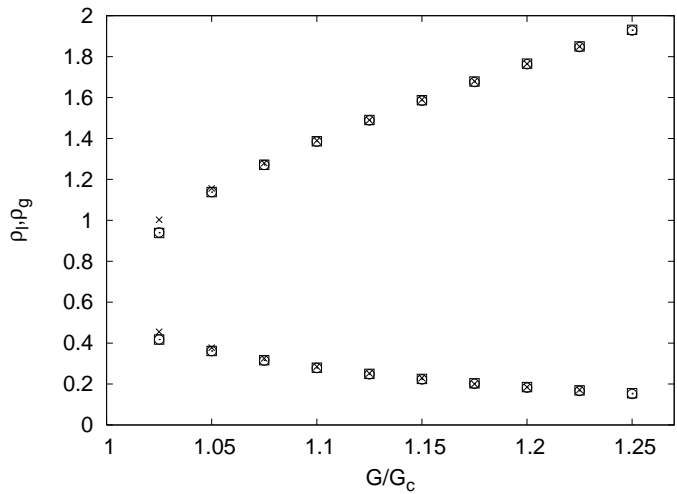}
    \caption{Results obtained from the mechanical stability conditions
      (\ref{eq:inte}), $\square$, and thermodynamic coexistence curve
      (\ref{eq:max}), $\circ$, for liquid and gas densities.  The
      results of numerical simulation are also reported
      ($\times$). The numerical simulations are carried out in a fully
      periodic $2d$ setup with grid mesh $L_{x}\times L_{y}=90\times
      90$ and a relaxation time $\tau=1$. The system is then initiated
      to have a flat interface along the $x$ direction. The top branch
      refer to the liquid density, $\rho_l$, while the bottom branch to
      the gas density, $\rho_g$. }
    \label{fig:comp}
  \end{center}
\end{figure}

In this section, we want to discuss a few issues on the thermodynamics
consistency of the SC model here studied. Let us first start from the
simple case of a liquid-gas interface in absence of boundaries.  Up to
now, in order to get the whole density profile, we have used the
mechanical equilibrium condition supplied with the appropriate
boundary conditions for the densities in the two phases, see
eqs. (\ref{eq:density}--\ref{eq:inte}). Next, we investigate the
relation between this mechanical condition and the thermodynamic
conditions on both the density values and on the density profile
(i.e. on the surface tension expression).  The Maxwell construction
which determines the thermodynamics consistency (see also
\cite{Callen}) in the $(P,V)$ diagram of the liquid-gas interfaces is
built in terms of the requirement that: $\int_{\rho_g}^{\rho_l}
[P_{bulk}-P_b(\rho)]dV =0$ where $V \propto 1/\rho$.  It is easy to realize that
this condition leads to the following integral constraint for the two
densities, $\rho_l,\rho_g$, at phase coexistence: 

\be
\label{eq:max}
\int_{\rho_l}^{\rho_g} \left[P_{bulk}-c_s^2 \rho
  -\frac{c_s^2\G_b}{2}\psi^{2}(\rho)\right] \frac{1}{\rho^{2}} d \rho
=0 \ee
which coincides with (\ref{eq:inte}) only if $\psi(\rho) \propto
\rho$.  This is the indication that the SC choice $\psi(\rho) =
(1-\exp\left(-\rho/\rho_{0}\right))$ is thermodynamically inconsistent,
although one may notice that the discrepancy are extremely small in
all realistic cases. For instance in figure \ref{fig:comp} we show the
good agreement of the liquid and gas density values obtained by using
the mechanical equilibrium equation (\ref{eq:inte}) and the Maxwell
construction (\ref{eq:max}) for different temperatures ($\G_{b}$).

Another problem using the SC approach to describe the interface shape
consists in the expression for the surface tension (\ref{tensione})
which is proportional to $(\partial_y \psi)^2$ instead of being
proportional to $(\partial_y \rho)^2$ as required by thermodynamical
arguments (see \cite{rw}). Also in this case, however, the situation
is quite encouraging. In fact, let us notice that starting from
(\ref{eq:psi1}) and the expression of the bulk pressure
(\ref{eq:eqstate}), the critical point of the system is identified by
the relations: 

\be\label{1derivative} \frac{\partial P_{b}(\rho)}{\partial \rho}=0;
\qquad
%\ee
%\be\label{2derivative}
\frac{\partial^{2} P_{b}(\rho)}{\partial \rho^{2}}=0.  \ee 

These are equivalent to

\be
\label{1d1}
(\psi^{\prime})^{2}=-\psi \psi^{\prime \prime}; \qquad
%\ee
%\be\label{1d2}
\psi \psi^{\prime}=-\frac{1}{\G_{b}}.
\ee
Using the functional form $\psi(\rho)=(1-\exp(-\rho/\rho_{0}))$ it is
readily checked that: 
\be
\label{1id0} \psi \psi^{\prime \prime}=-\frac{1}{\rho_{0}}\psi \psi^{\prime}.  
\ee 
Now, using
(\ref{1derivative},\ref{1d1},\ref{1id0}) we obtain 
\be
(\psi^{\prime})^{2}=-\psi \psi^{\prime \prime}=\frac{1}{\rho_{0}} \psi
\psi^{\prime}=-\frac{1}{\rho_{0} \G_{b}}.  
\ee
 Since $\psi^{\prime}=\partial \psi/\partial \rho$, this suggests the
following scaling relation

\be \label{eq:n} 
|\partial_{y} \psi|^{2} \sim -\frac{1}{\G_{b} \rho_{0}}
|\partial_{y} \rho|^2
\ee
which in turn would imply that the correct matching

\be \gamma_{lg}= -\frac{1}{2}c^{4}_{s}\G_{b}
\displaystyle\int_{-\infty}^{+\infty}|\partial_{y}\psi|^{2} dy \sim
\frac{c^{4}_{s}}{2 \rho_{0}}
\displaystyle\int_{-\infty}^{+\infty}|\partial_{y}\rho|^{2} dy .
\label{eq:sftrue}
\ee 

\begin{figure}
  \begin{center}
    \includegraphics[scale=.75]{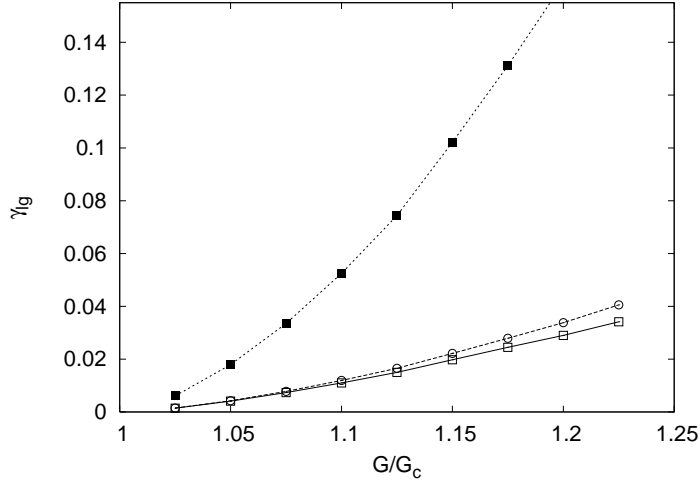}
    \caption{Surface tension as a function of the interaction
      parameter $\G_{b}$. The surface tension calculated  using both
      expressions in equation (\ref{eq:sftrue}), the case with
      $\partial_y \rho$ ($\circ$) and the case with $\partial_y \psi$
      ($\square$). To show the importance of the normalization factor
      (\ref{eq:n}) we also show the surface tension calculated by
      simply replacing $\partial_y \psi \to \partial_y \rho$ in
      (\ref{tensione}), (filled squares).}
    \label{fig:comparisonSF}
  \end{center}
\end{figure}

The previous argument, although exact at the critical point
($\G_{b}=\G_{c}$), is semi-quantitatively correct for $|\G_{b}|>|\G_{c}|$.
In fact, in figure \ref{fig:comparisonSF} we show the comparison
between the surface tension measured in our LBE approach and the one
defined by (\ref{eq:sftrue}). The agreement is quite satisfactory for
all values of temperature, showing that the model is not far from
being consistent also on that side.

\subsection{Two-phase mesoscopic model  vs. Navier-Stokes equations}
In this section we discuss the interplay between the mesoscopic
two-phase approach here proposed in presence of boundaries and the
most traditional ``fully macroscopic'' description at the
hydrodynamical level with ad-hoc boundary conditions. A similar study
has already been proposed by \cite{MD8} where a comparison between
microscopic MD and macroscopic Stokes flows with slip boundary
conditions was presented.  In our case, we focus on two analytical
results obtained for Navier-Stokes equations in a channel with
longitudinal or transverse (with respect to the flow direction)
free-slip strips, i.e. the case when some inhomogeneous material is
deposited at the surface so as to drastically change the boundary
conditions of the Navier-Stokes field from no-slip, ${\bm u}_{||}=0$,
to free-slip, $\partial_{n} {\bm u}_{||}=0$. Here ${\bm u}_{||}$
stands for the velocity component along the surface and $\partial_{n}$
is the derivative along the surface normal direction.

This problem can be attacked at a purely hydrodynamical level,
bypassing completely the chemical-physical reactions at the wall which
generates these two different boundary conditions and focusing only on
the bulk liquid incompressible phase. Using conformal mapping on plane
surfaces made up of longitudinal strips, (\cite{Phi1,Phi2})
proposed an exact analytical solution for the Stokes problem and only
very recently a similar calculation has been proposed for transverse
strips (\cite{stone}). In a previous paper we have shown that the same
analytical results can be obtained within the realm of LBE for single
flows, but with properly modified boundary conditions
(\cite{bbsst05}). The motivations were very close to the Navier-Stokes
continuum approach: neglect microscopic details very close to the
boundary, including the possible presence of a rarefaction layer and
try to mimic the bulk profile by {\it renormalizing} the fluid-wall
interactions into a suitable {\it wall function}. Next we show that
indeed one can recover both the analytical results of \cite{stone} and
\cite{Phi1,Phi2} and the numeric of the single phase LBE, by using the
present two-phase model. In this way, we fill the gap between the bulk
physics and the boundary layer physics in the proximity of a
chemically active wall and we are able to follow also the dynamics
inside this latter layer. To accomplish this goal, we performed LBE
simulations of the SC model in a channel where $\psi(\rho_w)$ had a
periodic structure with alternating strips of hydrophobic
($\rho_{w}=\rho_{g}$, local contact angle $180^{\circ}$) and
hydrophilic ($\rho_{w}=\rho_{l}$, local contact angle $0^{\circ}$)
materials. In figure \ref{fig:stonelauga} (left panel) we show the
analytical results for the effective slip length $\lambda_s$ obtained
(for both longitudinal and transverse strips) by means of the NS
approach and by the present LBE-SC method.  The agreement between the
two is very good. In the right panel, in order to capture the meaning
of local boundary conditions obtained using alternating patterns of
wetting and non wetting strips, we compare our multiphase mesoscopic
approach with the integration of the incompressible lattice Boltzmann
Equation with alternating strips of no-slip and free shear
(\cite{bbsst05}). As we can see, while the non wetting strips simply
reduce to a parabolic Poiseuille profile, the presence of the non
wetting strips acts as a free shear zone where, due to the presence of
the gas, the liquid can slide away producing the local free-shear
condition.

\begin{figure}
  \begin{center}
    \includegraphics[scale=0.3]{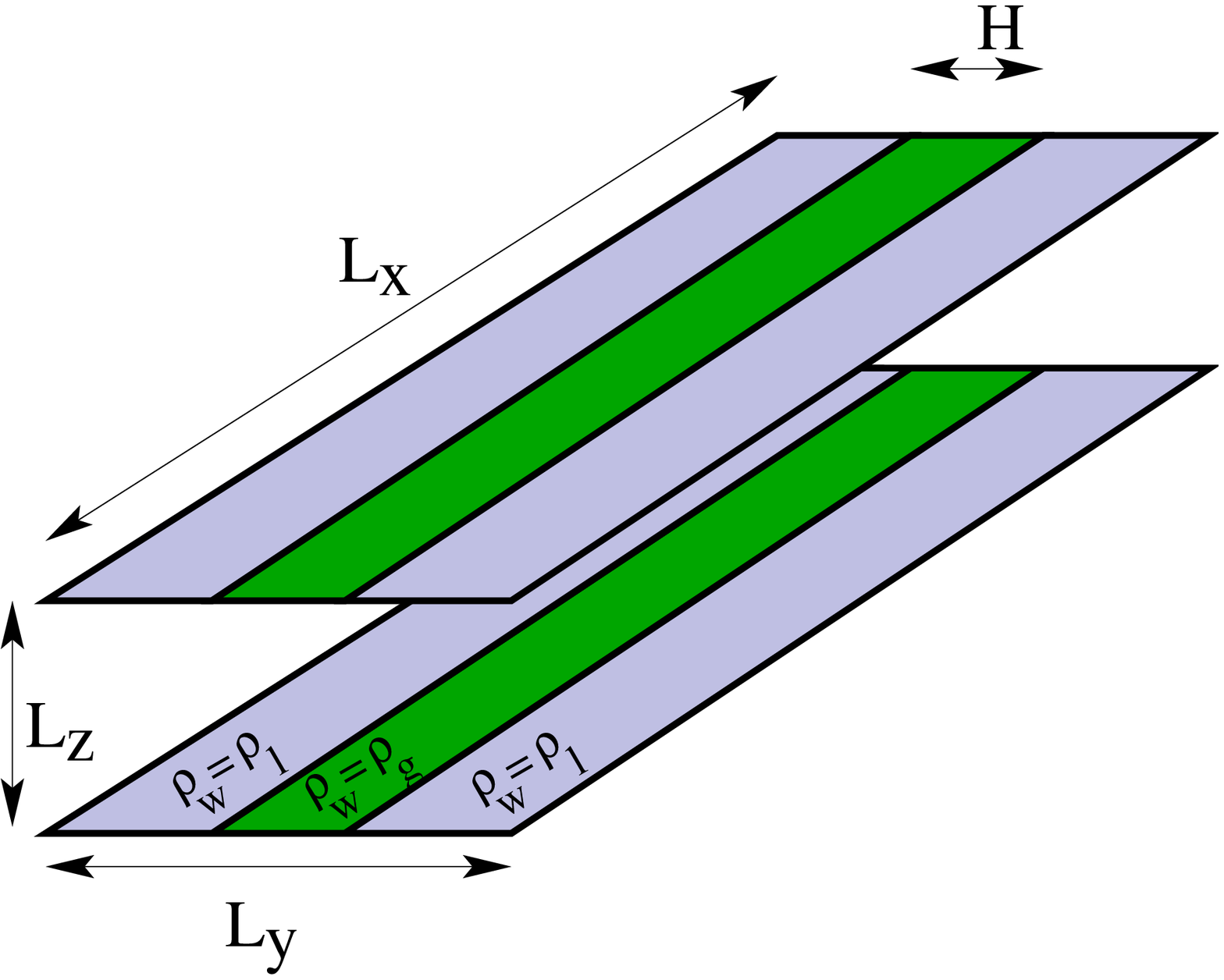}
    \includegraphics[scale=0.3]{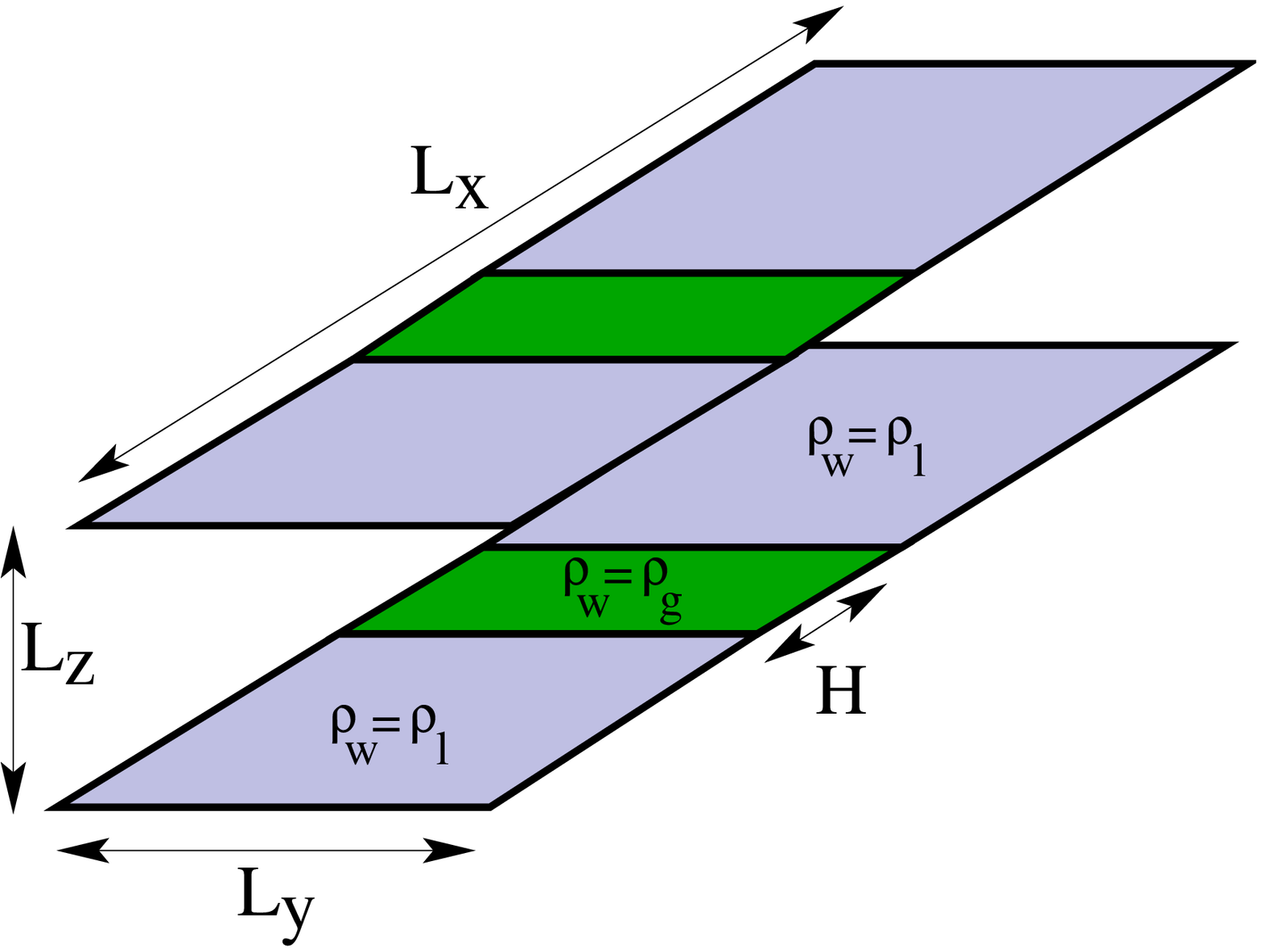}
    \caption{Schematic configurations for the simulations of laminar
      flows with boundary conditions of free shear and no slip. The
      alternating pattern of longitudinal (left) and transverse
      (right) strips of free shear is produced by varying the boundary
      density from complete wetting ($\rho_{w}=\rho_l$, contact angle
      $0^{\circ}$, no slip) and perfect dewetting in a strip $H$
      ($\rho_{w}=\rho_{g}$, contact angle $180^{\circ}$, free
      shear). With these setups the 'slip' probability is $\xi=H/L$
      being $L=L_{x} (L_{y})$ for transverse (longitudinal) strips.
      All the details of numerical simulations are give in figure
      \ref{fig:stonelauga}.}
    \label{fig:stonelaugasketch}
  \end{center}
\end{figure}

\begin{figure}
  \begin{center}
    \includegraphics[scale=0.55]{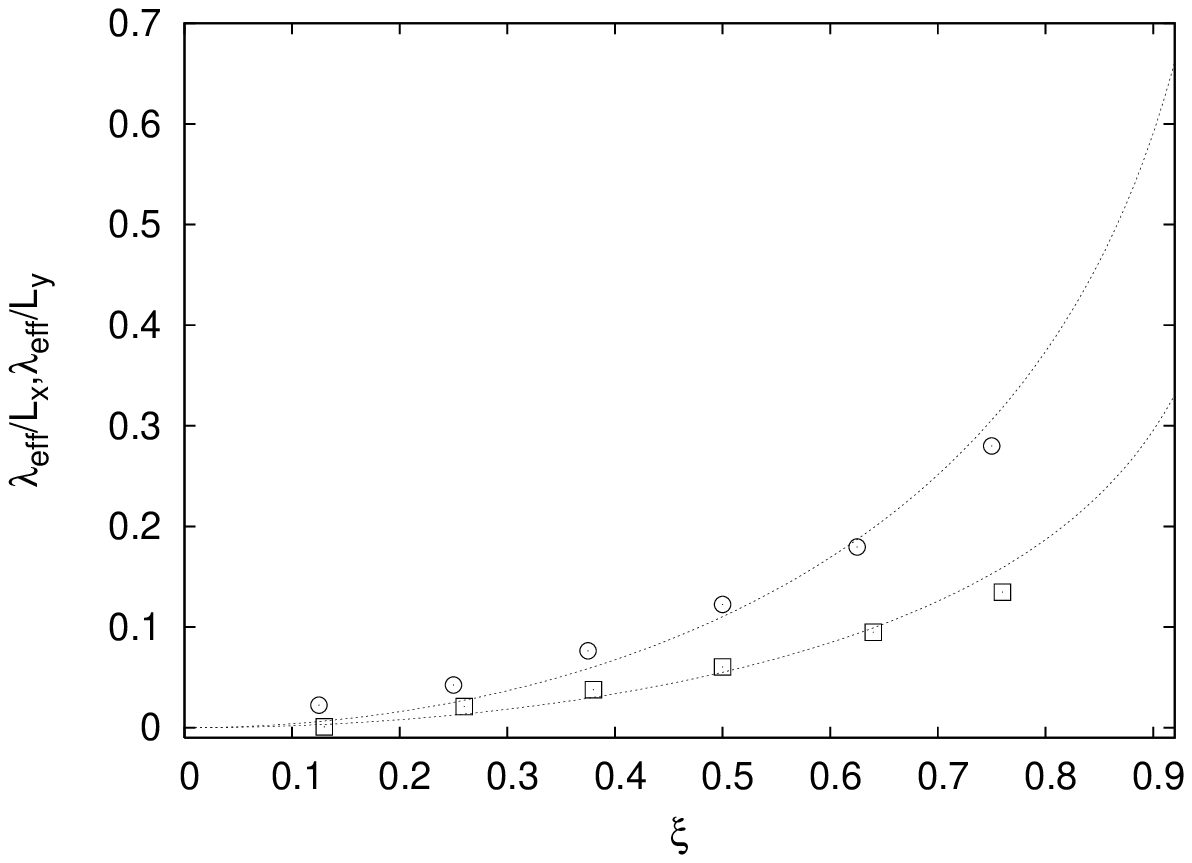}
    \includegraphics[scale=0.55]{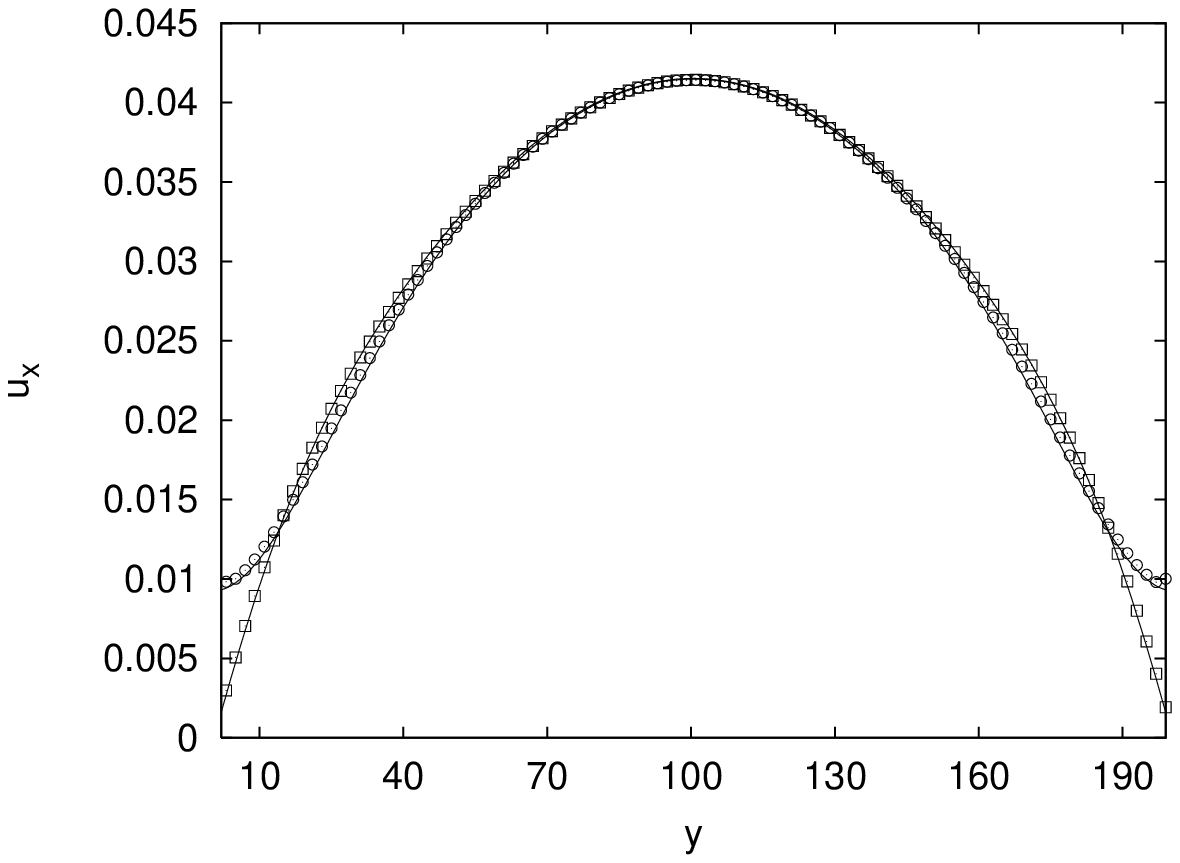}
    \caption{Effective slip lenght as a function of the free shear
      percentage ($\xi$) in multiphase approach with alternating
      strips of wetting and non-wetting properties. The numerical
      simulations are carried out in $3D$ setups with $L_{x}\times
      L_{y} \times L_{z}=64\times 64\times 200$ with periodic boundary
      conditions in the stream-wise ($x$) and span-wise directions
     ($y$). The interaction parameter is $\G_{b}=-6.0$ and the 
      relaxation time is $\tau=1$. The corresponding liquid and gas
      densities ($\rho_{l}=2.65,\rho_{g}=0.07$) are used to mimic no
      slip and free shear respectively. Then, the steady state
      effective slip lengths for transverse ($\square$) and
      longitudinal ($\circ$) strips normalized to the pattern
      dimension ($L_{x}$ for transverse and $L_{y}$ for longitudinal)
      are compared with exact analytical estimates given for stokes
      flow with alternating non slip and free shear (lines). Right
      panel: The velocity profile in the inlet of the channel and in
      the middle of a free shear strip from the integration of the
      incompressible lattice Boltzmann equation with mixed boundary
      conditions (straight lines) are compared with the results of the
      mesoscopic multiphase approach ($\square$ in the inlet and
      $\circ$ in the middle of free shear strip). The incompressible
      Lattice Boltzmann equation has been used with the same relaxation
      time and an average density equal to the one of the mesoscopic
      approach. Both simulation have been forced with a constant
      pressure gradient so as to reproduce a center channel velocity
      equal to $u_{c}=0.04$ in LB units.}
\label{fig:stonelauga}
\end{center}
\end{figure}

\subsection{Interaction with the wall: the role of the density gradient}
Up to now we have modeled the wall properties using a single
parameter, $\psi(\rho_w)$, which fixes the density of the flow at the
boundary. These parameters can be intended as the counterpart of the
interaction energy between solid-liquid molecules in an atomistic
approach. In principle, also the typical interaction length may play a
crucial role, the existence of a characteristic distance is the result
of the interplay between attractive and repulsive terms in the
Lennard-Jones potential. In order to mimic this effect, it has been
proposed to enrich the mesoscopic LBE by assuming that the interaction
with the wall is supplemented by another external force $F_w$ normal
to the wall and decaying exponentially (\cite{KWOK2,sullivan}):
\be\label{wallint} F_{w}(\x)=\G_{w} \rho(\x) e^{-
  \left|\x-\x_{w}\right| / \eta} \ee
where $\x_{w}$ is a vector running along the wall location and $\eta$
the typical length-scale of the fluid-wall interaction, also known as
the Kac range parameter (\cite{sullivan}).  Equation (\ref{wallint}) has
been previously used in literature in connection with a slightly
different LBE scheme, to show numerically how the wetting angle depends on the
ratio $\G_{w}/\G_{b}$ in the presence of phase coexistence between
vapor and liquid (\cite{KWOK2}).

The introduction of an external force, exponentially damped in the
bulk of the flow, allows the model to control also the gradient of the
density profile at the wall. This is a two parameter model able to fit
with high accuracy the value of the density at the wall, through
$\rho_w$, and the derivative of the density at the wall, $\partial_y
\rho$, through the $\G_w$ term in (\ref{wallint}).

For the case with $\G_w $, one must modify the structure of the
pressure tensor as follows
\begin{eqnarray}
P_{ij}=\left[
  c^{2}_{s}\rho+\frac{1}{2}c^{2}_{s}\G_{b}\psi^{2}+\frac{1}{2}c^{4}_{s}\G_{b}\psi
  \Delta \psi +\frac{1}{4}\G_{b}c^{4}_{s} |{\bm \nabla} \psi|^{2}
  \right]\delta_{ij}\nonumber \\ -\frac{1}{2}c^{4}_{s}\G_{b}
\partial_{i}\ \psi \partial_{j} \psi + \delta_{iy} \G_w
\displaystyle\int_{0}^{y} \rho(s) e^{-s/\eta} ds.
\end{eqnarray}
The off diagonal term, $P_{xy}$, remains the same, while the mismatch
between the pressure tensor parallel to the interface, $P_{xx}$ and
the pressure term perpendicular to the interface, $P_{yy}$ is changed
to:

\be P_{xx}=P_{yy}-\frac{c^{4}_{s}}{4} \G_b \psi \partial_{y}
\partial_{y} \psi-\G_w \displaystyle\int_{0}^{y} \rho(s) e^{-s/\eta}
ds .\ee This implies that the previous estimate (\ref{eq:ca}) of the
contact angle must be replaced by:

\be \cos{(\theta)}=\frac{\displaystyle\int_{sg} \left( |\partial_{y}
  \psi|^2 - \frac{2 \G_w}{{\cal G}_{b} c_s^4}
  \displaystyle\int_{0}^{y} \rho(s) e^{-s/\eta} ds \right)
  dy-\displaystyle\int_{sl} \left( |\partial_{y} \psi|^2-\frac{2
    \G_w}{{\cal G}_{b} c_s^4} \displaystyle\int_{0}^{y} \rho(s)
  e^{-s/\eta} ds \right) dy}{\displaystyle\int_{lg} |\partial_{y}
  \psi|^2 dy}.
\label{eq:cagwall}
\ee 

\begin{figure}
  \includegraphics[scale=.6]{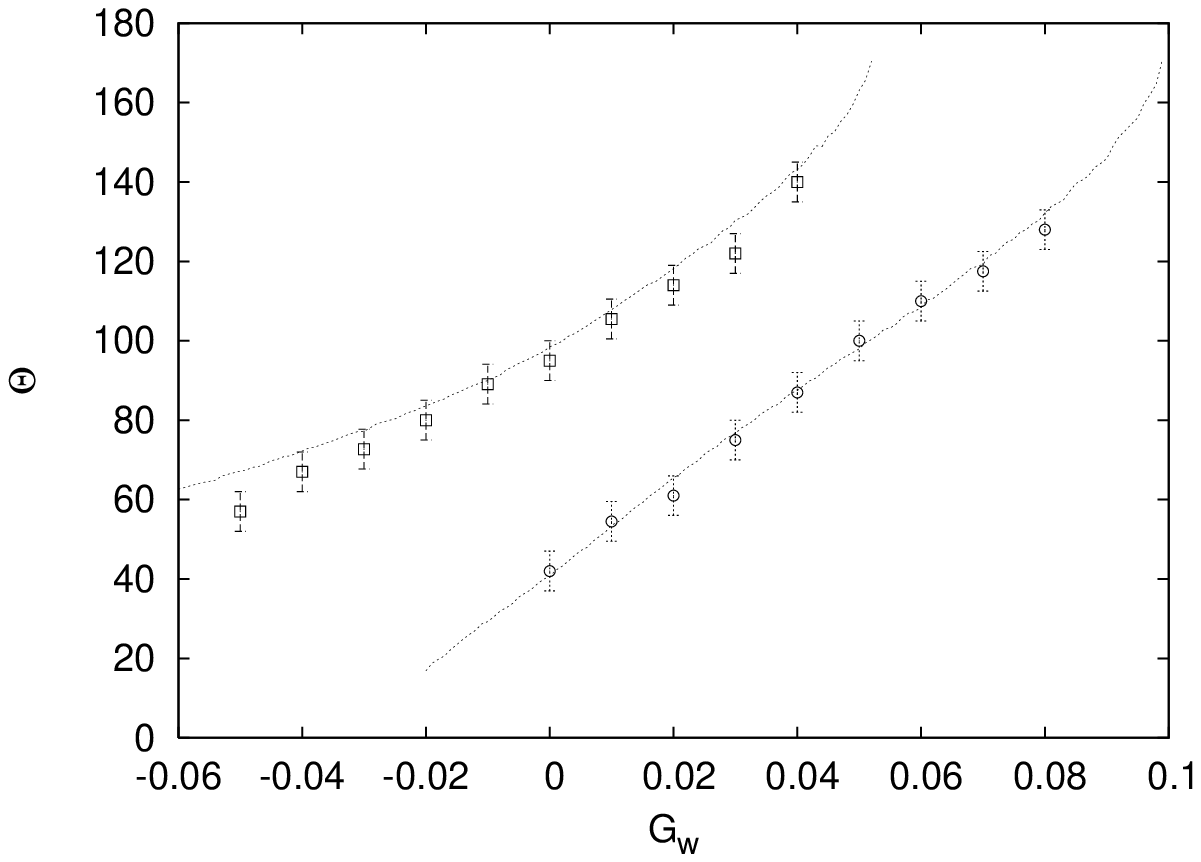}
  \includegraphics[scale=.6]{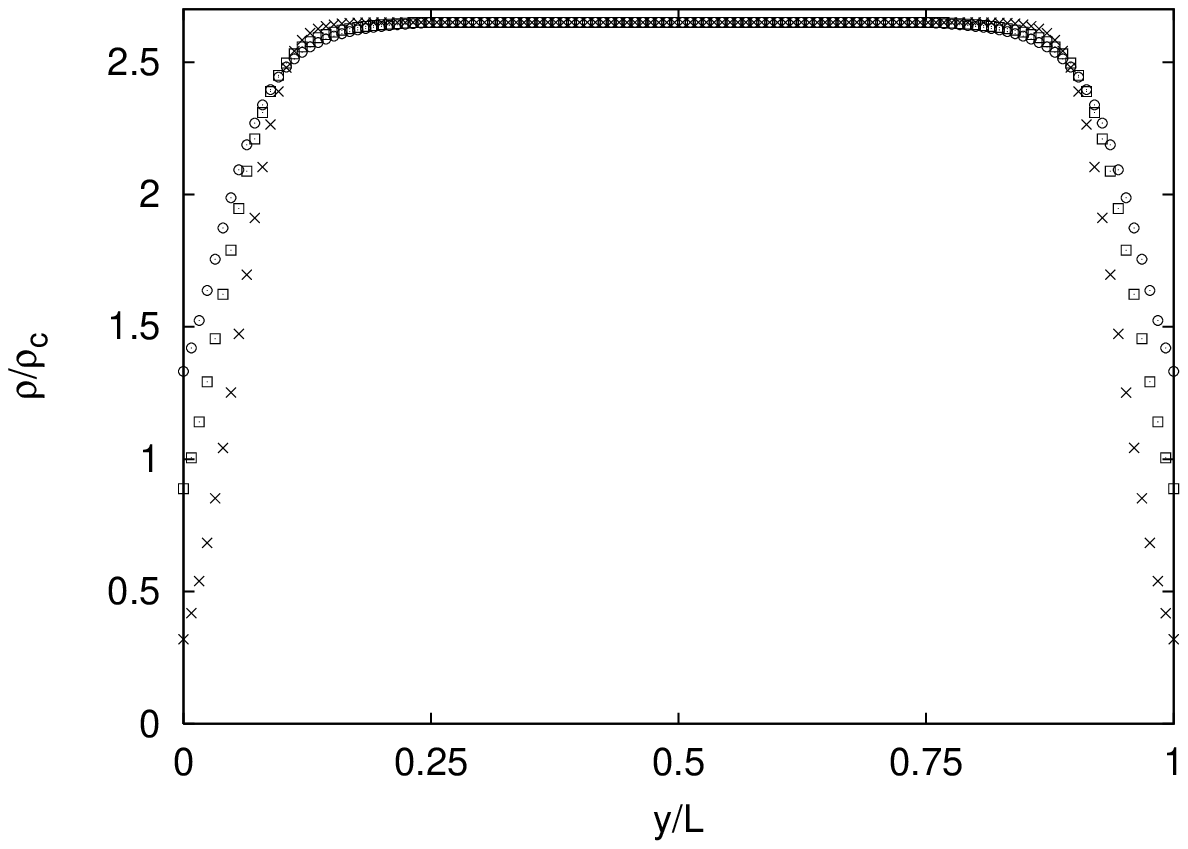}
  \caption{Contact angle for a two parameter model. Left panel: for a
    given value of $\rho_{w}$ we show the contact angle for different
    values of $\G_{w}$: $\rho_{w}=1.3$ ($\circ$), $\rho_{w}=0.6$
    ($\square$). The analytical estimate using the mechanical
    stability is compared with lattice Boltzmann simulations. The
    details of the numerical simulations are the same of figure
    \ref{figcontact} with the introduction of a wall-fluid force as in
    (\ref{wallint}). Left panel: for the same contact angle
    ($130^{\circ}$) we report different profiles obtained for
    different choices of the parameters ($\rho_{w},\G_{w}$). In
    particular the value of $\rho_w$ chosen is:
    $\rho_{w}=1.3$($\circ$) $\rho_{w}=0.85$($\square$)
    $\rho_w=0.3$($\times$) and the corresponding value of ${\cal
      G}_{w}$ has been chosen to reproduce the same contact angle of
    $130^{\circ}$ estimated from (\ref{eq:cagwall}).}
  \label{fig:gwall}
\end{figure}

The physics at the boundary now depends on two parameters, which may
change the contact angle and the density profile
independently. Unfortunately it is very difficult to make any
quantitative measurement of the density profile in experiments or MD
simulations.  It is therefore difficult to asses in a systematic way
the potential of the LBE model.  In figure \ref{fig:gwall} we show the
contact angle that can be measured and calculated using
(\ref{eq:cagwall}) in our LBE approach. In the same figure (right
panel) we also show the variation in the density profile for a few
values of the couple $(\G_w,\rho_w)$ leading to the same contact
angle.
  As one can see, the model is
very sensitive to different choices of the two free parameters, which
makes it potentially useful to describe physical situations with
large variations in the density profiles in the proximity of the
boundaries.

\section{Conclusions and perspectives}
We have presented a mesoscopic model, based on the Boltzmann
Equations, for the interaction between a solid wall and a non-ideal
fluid.  The model is an extension of the SC model for dense fluids in
unbounded domains. We have first derived an analytical expression for
the {\it contact angle} and for the surface energy between any two of
the liquid, solid and vapor phase, by introducing a parameter,
$\psi(\rho_w)$ which fixes the density value of the Boltzmann
molecules at the solid wall.  We have shown how in this way one can
cover the whole range of contact angles, from a super-hydrophobic
surface $\theta \sim 180^{\circ}$ to a condition of perfect
wettability $\theta \sim 0^o$. Concerning the thermodynamic
consistency of the model we have shown that although not formally
verified it does not introduce any systematic error in the results for
the surface tension and for the liquid-gas density variations at
changing the system temperature (Maxwell construction). We have
discussed the connection between the rarefaction layer in the
proximity of the wall and the production of an {\it apparent slip}
phenomenon.  We have also presented a comparison between the results
obtained within the realm of our two-phase mesoscopic model and some
analytical expression for the slip length calculated within a
Navier-Stokes approach of a single phase fluid with alternating
boundary conditions formed of free shear and no-slip strips. We have
shown that the LBE approach is able to reproduce {\it quantitatively}
the analytical results, supporting the statement that slip phenomena
at the macroscopic Navier-Stokes level can be interpreted by an
apparent slip induced by gas accumulation close to the boundary.\\ We
have also studied the effects on the contact angle and on the density
profiles of the use of a second free parameter, $\G_w$ connected to
the wall-force decaying exponentially inside the bulk of the
flow. This second parameter plays an important role if one wants to
control both the value of the density at the wall and its gradient.\\ An
important outcome of this study is the possibility to fully integrate
the two-phase approach with complex geometries (see also
\cite{BE2phase5}). For instance, recent MD studies \cite{barrat} have
demonstrated the existence of a wetting/dewetting transition in
micro-channel with grooves. The effect is driven by capillarity forces
which may expel the liquid out of the corrugation leading to an
increase of the {\it effective} slip length and of the mass flow
rate. The mesoscopic model here presented is fully capable to
reproduce the same effects. Moreover, it may also implement more
complex surface patterns (different corrugations) then those possible
with MD simulations maintaining a control on the spatial and temporal
fluctuations at the macroscopic scale.  Results on this direction will
be reported elsewhere (\cite{NatMat}).

We acknowledge useful discussion with D. Lohse who suggested to
perform the comparison with the Navier-Stokes equations in the case of
longitudinal and transverse free-slip strips. We also thank X. Shan
for stimulating discussion.

\section{Appendix A}
In this appendix we detail the calculation needed to perform the
continuum limit for the nonideal pressure term. Let us first notice
that the value of $\Delta t$ just enters the discretized LBE equation
(\ref{eq:lbe}) as a normalizing factor with respect to the relaxation
time $\tau$. Changing $\Delta t$ just means to redefine $\tau$.

Now, if we start from the expression for the $i$ component of the
interparticle force: 
\be
\label{FORCE} 
F_{i}({\bm x},t) = -\G_{b} \psi({\bm x})\sum_\alpha
w_{\alpha} \psi({\bm x}+{\bm c_{\alpha}} \Delta t,t)c^{i}_{\alpha} \ee
assuming a stationary state (${\bm F}({\bm x},t)={\bm F}({\bm x})$)
and Taylor expanding  up to the third order we obtain: 
\begin{eqnarray}
 F_{i}({\bm x})&=&-\G_{b}\psi({\bm x})[\sum_{\alpha} w_{\alpha} 
c^{i}_{\alpha} \psi({\bm x}) + \Delta t \sum_{\alpha,j} w_{\alpha} 
c^{i}_{\alpha} c^{j}_{\alpha} \partial_{j} \psi({\bm x}) + \nonumber  \\
& & \frac{(\Delta t)^2}{2} \sum_{\alpha,j,k} w_{\alpha} c^{i}_{\alpha} c^{j}_{\alpha} c^{k}_{\alpha}
\partial_{j} \partial_{k} \psi({\bm x})+\frac{(\Delta t)^3}{6}
\sum_{\alpha,j,k,l} w_{\alpha} c^{i}_{\alpha} c^{j}_{\alpha} c^{k}_{\alpha}
c^{l}_{\alpha} \partial_{j} \partial_{k} \partial_{l} \psi({\bm x})]. \nonumber  
\end{eqnarray}

If we choose the weights $w_{\alpha}$ to satisfy the following tensor
relations

$$\sum_{\alpha}w_{\alpha}c^{i}_{\alpha}=0$$
$$\sum_{\alpha}w_{\alpha} c_{\alpha}^{i}c_{\alpha}^{j}=c^{2}_{s} \delta_{ij}$$
$$\sum_{\alpha}w_{\alpha} c_{\alpha}^{i}c_{\alpha}^{j}c^{k}_{\alpha}=0 $$
$$\sum_{\alpha}w_{\alpha}
c_{\alpha}^{i}c_{\alpha}^{j}c_{\alpha}^{k}c_{\alpha}^{l}=c^{4}_{s}(\delta_{ij}
\delta_{kl} + \delta_{ik} \delta_{jl} + \delta_{il} \delta_{jk}) $$ we
end up with the following expression for the $i$-component of ${\bm
  F}$: \be F_{i}=\G_{b} \psi \frac{\Delta t}{2} c^{2}_{s} \partial_{i}\psi +
\G_{b} \psi \frac{(\Delta t)^3}{2} c^{4}_{s}\partial_{i}\Delta\psi.  \ee being
$\Delta$ the Laplacian operator. The above expression for the
interparticle force can be easily translated into an excess pressure
with respect to the ideal gas expression ($c^{2}_{s}\rho$) using the
definition: \be -\partial_{j} P_{i,j} -\partial_{i}(c^{2}_{s}\rho) =
F_{i} \ee 
and therefore we end up with a pressure tensor of the form:
\be
\label{TENSORE1}
 P_{ij}=\left[ c^{2}_{s}\rho+\frac{\Delta
     t}{2}c^{2}_{s}\G_{b}\psi^{2}+\frac{(\Delta
     t)^3}{2}c^{4}_{s}\G_{b}\psi \Delta \psi + \frac{\G_{b}c^{4}_{s}
     (\Delta t)^3}{4} |{\bm \nabla} \psi|^{2} \right]
 \delta_{ij}-\frac{(\Delta t)^3}{2}c^{4}_{s}\G_{b}\partial_{i} \psi
 \partial_{j}\psi.  \ee
An important remark is now in order. The continuum expression just
derived depends explicitly on the interparticle potential range, here
of the order of $c_s \Delta t$. The limit $\Delta t \rightarrow 0$
would therefore imply a vanishing interaction range, i.e. the ideal
gas limit, $P_{ij} = c_s^2 \rho \delta_{ij}$.

\section{Appendix B}
\begin{figure}
  \begin{center}
    \includegraphics[scale=0.45]{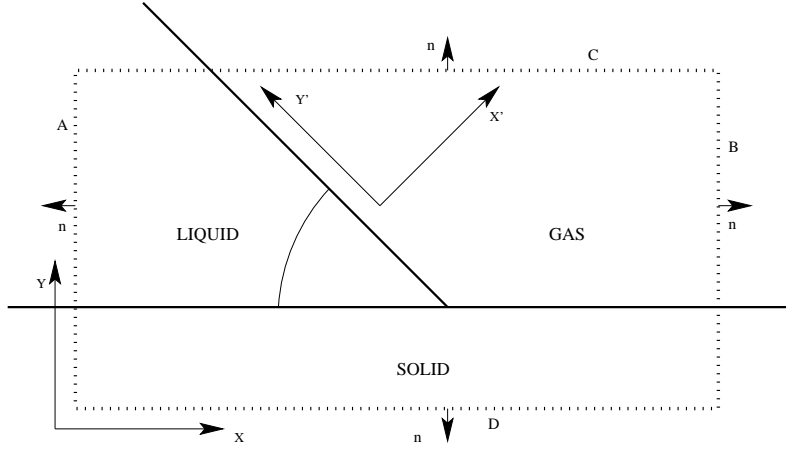}
    \caption{The schematic figure of the contact angle setup.}
    %\label{fig:angle}
    \label{fig:angle}
  \end{center}
\end{figure}

In this appendix we will detail the calculation of the contact angle
used throughout the text. Starting from the pressure tensor (for $\Delta t=1$) 

\be P_{ij}=\left[
  c^{2}_{s}\rho+\frac{1}{2}c^{2}_{s}\G_{b}\psi^{2}+\frac{1}{2}c^{4}_{s}\G_{b}\psi
  \Delta \psi +\frac{1}{4}\G_{b}c^{4}_{s} |{\bm \nabla} \psi|^{2}
  \right]\delta_{ij}-\frac{1}{2}c^{4}_{s}\G_{b} \partial_{i}\psi
\partial_{j}\psi .  \ee in the coordinate system of figure
(\ref{fig:angle}) we make the following change of variables \be \left
\{ \begin{array} {l} x^{\prime}=+x \sin(\theta) + y
  \cos(\theta)\\ y^{\prime}=-x \cos(\theta) + y \sin(\theta)
\end{array} \right.
\ee 

being $y^{\prime}$ aligned with the separation line between liquid and
gas.

\be \left \{ \begin{array} {l} \frac{\partial }{\partial
    x}=\frac{\partial x^{\prime}}{\partial x} \frac{\partial}{\partial
    x^{\prime}}+\frac{\partial y^{\prime}}{\partial x} \frac{\partial
  }{\partial y^{\prime}}=\sin(\theta) \frac{\partial}{\partial
    x^{\prime}} - \cos(\theta)\frac{\partial}{\partial
    y^{\prime}}\\ \frac{\partial}{\partial y}=\frac{\partial
    x^{\prime}}{\partial y} \frac{\partial}{\partial
    x^{\prime}}+\frac{\partial y^{\prime}}{\partial y}
  \frac{\partial}{\partial y^{\prime}}=\cos(\theta)
  \frac{\partial}{\partial x^{\prime}}+\sin(\theta)
  \frac{\partial}{\partial y^{\prime}}
\end{array} \right.
\ee

If we assume only the $x^{\prime}$ dependence we can write

\be \psi(x,y)=\psi(x^{\prime}) \ee \be \left \{ \begin{array} {l}
  \frac{\partial}{\partial x}= \sin{(\theta)}\frac{\partial}{\partial
    x^{\prime}} \\ \frac{\partial}{\partial y}= \cos{(\theta)}
  \frac{\partial }{\partial x^{\prime}}
\end{array} \right.
\ee \be \displaystyle\int dx \rightarrow \frac{1}{\sin{(\theta)}}
\displaystyle\int d x ^{\prime} \ee which clearly imply \be dx=\sin{(
  \theta)} d x^{\prime} \ee and \be\label{MIX} \displaystyle\int
(\partial_{x} \psi) (\partial_{y} \psi) dx=\frac{1}{\sin{(\theta)}}
\displaystyle\int d x^{\prime} \sin{(\theta)} \cos{(\theta)}
(\partial_{x^{\prime}} \psi) (\partial_{x^{\prime}}
\psi)=\displaystyle\int d x^{\prime} \cos{(\theta)} (\partial_{x^{\prime}}
\psi) (\partial_{x^{\prime}} \psi).  \ee

By imposing a zero flux out of the boundaries of the vector
$(P_{xx},P_{xy})$ we obtain the mechanical definition of the contact
angle. To this purpose, let us notice that based on the definition of
the pressure tensor, we can write:

\be\label{relation1} P_{xx}=P_{yy}+\frac{c^{4}_{s}}{2} \G_{b}
\partial_{y} \psi \partial_{y} \psi \ee \be\label{relation2}
P_{xy}=\frac{c^{4}_{s}}{2} \G_{b} \psi \partial_{x} \partial_{y} \psi
\ee 

with $P_{yy}$ constant along the solid-liquid and solid-gas interface
(segments $A,B$ of figure \ref{fig:angle}). This, together with
(\ref{MIX}) implies

\be -\displaystyle\int_{sl} \frac{c^{4}_{s}}{2} \G_{b} \partial_{y}
\psi \partial_{y} \psi dy + \displaystyle\int_{sg} \frac{c^{4}_{s}}{2}
\G_{b} \partial_{y} \psi \partial_{y} \psi dy -\cos{(\theta)}
\displaystyle\int_{lg} \frac{c^{4}_{s}}{2} \G_{b}
\partial_{x^{\prime}} \psi \partial_{x^{\prime}} \psi dx^{\prime}=0
\ee and \be \cos{(\theta)}=\frac{-\displaystyle\int_{sl} |\partial_{y}
  \psi|^2 dy+\displaystyle\int_{sg} |\partial_{y} \psi|^2
  dy}{\displaystyle\int_{lg} |\partial_{x^{\prime}} \psi|^2
  dx^{\prime}}.  \ee
For the case with ${\cal G}_w$, the previously used pressure tensor
must be slightly modified due to the presence of a normal (say along
the $y$ direction) force between the fluid and the wall:
\be
P_{ij}=\left[
  c^{2}_{s}\rho+\frac{1}{2}c^{2}_{s}\G_{b}\psi^{2}+\frac{1}{2}c^{4}_{s}\G_{b}\psi
  \Delta \psi +\frac{1}{4}\G_{b}c^{4}_{s} |{\bm \nabla} \psi|^{2}
  \right]\delta_{ij}-\frac{1}{2}c^{4}_{s}\G_{b} \partial_{i}\ \psi
\partial_{j} \psi + \delta_{iy} \G_w \displaystyle\int_{0}^{y} \rho(s)
e^{-s/\eta} ds.  
\ee 
Then relations (\ref{relation1}) and (\ref{relation2}) should be
slightly modified since $P_{xy}$ gives the same result but for
$P_{xx}$ we have
\be P_{xx}=P_{yy}-\frac{c^{4}_{s}}{2} \G_b \partial_{y} \psi
\partial_{y} \psi - G_w \displaystyle\int_{0}^{y} \rho(s) e^{-s/\eta}ds 
\ee 
that imply the following estimate for the contact angle: 
\be
\cos{(\theta)}=\frac{-\displaystyle\int_{sl} \left( |\partial_{y}
  \psi|^2-\frac{2 G_w}{{\cal G}_{b} c_s^4} \displaystyle\int_{0}^{y}
  \rho(s) e^{-s/\eta} ds \right) dy+\displaystyle\int_{sg} \left(
  |\partial_{y} \psi|^2 - \frac{2 G_w}{{\cal G}_{b} c_s^4}
  \displaystyle\int_{0}^{y} \rho(s) e^{-s/\eta} ds \right)
  dy}{\displaystyle\int_{lg} |\partial_{x^{\prime}} \psi|^2
  dx^{\prime}} \ee
Notice that the importance of wall effects appears only in relation to
bulk terms in $\frac{2 G_w}{{\cal G}_{b} c_s^4}$.

\end{document}